\documentclass[aps,reprint,twocolumn,footinbib,showpacs,superscriptaddress,longbibliography]{revtex4-1}
\usepackage{graphicx}
\usepackage{dcolumn}
\usepackage{bm}
\usepackage{amssymb}
\usepackage{pifont}
\usepackage{amsmath}
\usepackage{times}

\begin{document}

\title{Macroscopic description for networks of spiking neurons}
\author{Ernest Montbri\'o}
\affiliation{Center for Brain and Cognition. Department of Information and Communication Technologies,
Universitat Pompeu Fabra, 08018 Barcelona, Spain}
\author{Diego Paz\'o}
\affiliation{Instituto de F\'{\i}sica de Cantabria (IFCA), CSIC-Universidad de
Cantabria, 39005 Santander, Spain }
\author{Alex Roxin}
\affiliation{Centre de Recerca Matem\`atica, Campus de Bellaterra, 08193 
Bellaterra (Barcelona), Spain.}

\date{\today}

\begin{abstract}
A major goal of neuroscience, statistical physics and nonlinear dynamics 
is to understand how brain function arises from the collective dynamics of networks of spiking neurons. This challenge has been chiefly addressed through 
large-scale numerical simulations. Alternatively, researchers have formulated mean-field 
theories to gain insight into macroscopic states of large neuronal networks 
in terms of the collective firing activity of the neurons, or the firing rate.  
However, these theories have not succeeded in establishing an exact correspondence between 
the firing rate of the network  
and the underlying 
microscopic state of the spiking neurons. 
This has largely constrained the range of applicability of such macroscopic 
descriptions, particularly when trying to describe neuronal synchronization.   
Here we provide the derivation of a set of exact macroscopic equations for 
a network of spiking neurons. Our results reveal that the spike generation 
mechanism of individual neurons introduces an effective coupling between two 
biophysically relevant macroscopic quantities, the firing rate and the 
mean membrane potential, which together govern the evolution of the neuronal 
network. The resulting equations exactly describe all possible macroscopic 
dynamical states of the network, including states of synchronous spiking activity. 
Finally we show that the firing rate description is related, 
via a conformal map, with a low-dimensional description in terms of the 
Kuramoto order parameter, called Ott-Antonsen theory.    
We anticipate our results will be an important tool in investigating 
how large networks of spiking neurons self-organize in time to process and 
encode information in the brain.  
\end{abstract}

\maketitle

Processing and coding of information in the brain necessarily imply the 
coordinated activity of large ensembles of neurons. Within sensory regions of the cortex, 
many cells show similar responses to a given stimulus, indicating a high 
degree of neuronal redundancy at the local level. This suggests that information 
is encoded in the population response and hence can be captured via macroscopic 
measures of the network activity \cite{ALP06}.
Moreover, the collective behavior of large neuronal 
networks is particularly relevant given that current brain measurement 
techniques, such as electroencephalography (EEG) or functional 
magnetic resonance imaging  (fMRI), provide data which is 
necessarily averaged over the activity of a large 
number of neurons.
 
The macroscopic dynamics of neuronal ensembles has been extensively 
studied through computational models of large networks of recurrently coupled 
spiking neurons, including Hodgkin-Huxley-type conductance-based neurons 
\cite{HH52} as well as simplified neuron models, 
see e.g.~\cite{BRC+07,IE08,Izh07}.
In parallel, researchers have sought to develop 
statistical descriptions 
of neuronal networks, 
mainly in terms of a macroscopic observable that measures 
the mean rate at which neurons emit spikes, the \emph{firing rate}
\cite{WC72,Ama74,Fre75,Ger95,FM98,AB97i,Bru00,
DA01,GK02,SHS03,Coo05,RBH05,DJR+08,ET10,OB11,RM11,SOA13}. 
These descriptions, called firing-rate equations (FREs), 
have been proven to be extremely useful in understanding general computational 
principles underlying functions such as memory \cite{Hop84,MBT08}, 
visual processing \cite{BLS95,HS98,MRR07}, 
motor control \cite{Zha96} or decision making \cite{WW06}.

Despite these efforts, to date there is no exact
theory linking the dynamics of a large network of spiking 
neurons with that of 
the firing rate. Specifically, current macroscopic descriptions 
do not offer a precise correspondence between the 
microscopic dynamics of the individual neurons, e.g.~individual membrane 
potentials, and the firing rate dynamics of the neuronal network. 

Indeed, FREs are generally derived through heuristic arguments 
which rely on the underlying spiking activity of the neurons being 
asynchronous and hence uncorrelated. As such, firing rate descriptions 
are not sufficient to describe network states involving some degree of spike 
synchronization. 
Synchronization is, however, an ubiquitous phenomenon 
in the brain, and its potential role in neuronal computation is the subject 
of intense research \cite{GK02,Abe91,VLR01,ES01,War03,Fri05,Buz06,FA11}. 
Hence, the lack of firing rate descriptions for synchronous states   
limits the range of applicability of mean-field 
theories to investigate neuronal dynamics. 

Here we propose a method to derive the FREs
for networks of heterogeneous, all-to-all coupled quadratic integrate-and-fire 
(QIF) neurons, which is exact in the thermodynamic limit, 
i.e.~for large numbers of neurons.   
We consider an ansatz for the distribution of the neurons' 
membrane potentials that we denominate the \emph{Lorentzian ansatz} (LA). 
The LA solves the corresponding continuity equation exactly,
making the system amenable to theoretical analysis. 
Specifically, for particular distributions of the heterogeneity, 
the LA yields a system of two ordinary differential equations
for the firing rate and mean membrane 
potential of the neuronal population. These equations 
fully describe the macroscopic states of the network
---including synchronized states---, and represent the
first example of an exact firing-rate description for a network of
recurrently connected spiking neurons. 
We finally show how the LA  
transforms, via a conformal mapping, into the so-called 
Ott-Antonsen ansatz (OA) that is extensively used to 
investigate the low-dimensional dynamics of large populations of 
phase oscillators in terms of the Kuramoto order parameter \cite{OA08}.

\section{Model description}

Hodgkin-Huxley-type neuronal models can be 
broadly classified into two classes, 
according to the nature of their transition to spiking 
in response to an injected current \cite{Hod48,RE98}.  
Neuronal models with so-called \emph{Class I} excitability,
generate action potentials with arbitrarily low frequency, 
depending on the strength of the applied current. 
This occurs when a resting state disappears
through a saddle-node bifurcation. In contrast, in neurons 
with \emph{Class II} excitability the action potentials are generated 
with a finite frequency. This occurs when the resting state loses 
stability via a Hopf bifurcation. The QIF neuron 
is the canonical model for \emph{Class I} neurons, and thus 
generically describes their dynamics near the spiking threshold 
\cite{EK86,LRN+00,Izh07}.
Our aim here is to derive the FREs corresponding to a heterogeneous 
all-to-all coupled population of $N$ QIF neurons. 
The correspondence  is exact in the thermodynamic limit, 
i.e.~when $N\to\infty$ 
(this convergence has been recently studied in \cite{bc13}). 

The microscopic state of the population 
of QIF neurons is given by the membrane potentials 
$\{V_j\}_{j=1,\ldots,N}$, which obey the following
ordinary differential equations~\cite{Izh07}: 
\begin{equation}\label{qif} 
\dot V_j= V_j^2+ I_j, \quad \text{if } V_j \geq V_p, 
\text{then } V_j \leftarrow  V_r.  
\end{equation}
Here the overdot denotes the time derivative, and $I_j$ represents
an input current. Each time a neuron's membrane potential $V_j$ reaches 
the peak value $V_p$, the neuron emits a spike and 
its voltage is reset to the value $V_r$. In our analysis we 
consider the limit $V_{p}=-V_r \to \infty$. This resetting rule 
captures the spike reset as well as the refractoriness of the neuron. 
Without loss of generality we have rescaled the time and the voltage 
in \eqref{qif} to absorb any coefficients which would have appeared in 
the first two terms. The form for the input currents is:

\begin{equation} 
I_j= \eta_{j}+J s(t)+I(t),
\label{qif2}
\end{equation}
where the external input has a heterogeneous, quenched component 
$\eta_j$ as  well as a common time-varying component $I(t)$, and
the recurrent input is the synaptic weight $J$ times the mean synaptic 
activation $s(t)$, which is written:
\begin{equation}
s(t)= \frac{1}{N} \sum_{j=1}^N  \sum_{k \backslash t_j^k<t}
\int_{-\infty}^{t}dt' a_\tau(t-t')\delta (t'-t_j^{k}).
\label{s}
\end{equation}
Here, $t_j^k$ is the time of the $k$th spike of $j$th neuron, $\delta (t)$ 
is the Dirac delta function, and $a_\tau(t)$
is the normalized synaptic activation caused by a single pre-synaptic 
spike with time scale $\tau$, e.g.~$a_\tau(t)=e^{-t/\tau}/\tau$.  
 
\subsection{Continuous formulation} 
 
In the thermodynamic limit $N\to \infty$, we drop the 
indices in Eqs.~\eqref{qif}, \eqref{qif2} and denote $\rho(V \vert \eta, t) d V$ 
as the fraction of neurons with membrane potentials
between $V$ and $V+dV$, and 
parameter $\eta$ at time $t$. Accordingly, 
parameter $\eta$ becomes now a continuous 
random variable distributed according to a probability distribution 
function $g(\eta)$.
The total voltage density at time $t$ is then given by
$\int_{-\infty}^{\infty} \rho(V \vert \eta, t)~g(\eta)~ d\eta$.

The conservation of the number of neurons leads to the 
following continuity equation:
\begin{equation}
\partial_t \rho + \partial_V\left[  (V^2+\eta+J s +I) \rho \right]=0 ,
\label{cont}
\end{equation}
where we have explicitly included 
the velocity given by \eqref{qif} and \eqref{qif2}.

\section{Results}

The continuity equation \eqref{cont} without temporal forcing $I(t)=0$
has a trivial stationary solution. 
For each value of $\eta$, this solution has the 
form of a Lorentzian function: $\rho_0(V|\eta) \propto (V^2+\eta+J s)^{-1}$. 
Physically, the Lorentzian density means that firing neurons 
with the same $\eta$ value will be scattered 
with a density inversely proportional to their speed \eqref{qif}, 
i.e.~they will accumulate at slow places and thin out at 
fast places on the $V$ axis. In addition, 
for those $\eta$ values corresponding to quiescent neurons,
the density $\rho_0$ collapses at the rest state in the form
of a Dirac delta function.

Next we assume that, independently of
the initial condition, solutions of \eqref{cont} generically converge 
to a Lorentzian-shaped function,
so that all relevant dynamics occur inside that lower dimensional space.
This fact is mathematically
expressed by the following Lorentzian Ansatz (LA)
for the conditional density functions:
\begin{equation}
\rho(V \vert \eta, t)= \frac{1}{\pi}\frac{x(\eta,t)}{\left[V-y(\eta,t)\right]^2
+x(\eta,t)^2},
\label{la}
\end{equation}
which is a Lorentzian function with time-dependent half-width 
$x(\eta,t)$ and center at $y(\eta,t)$. In the following we assume 
that the LA \eqref{la} completely describes the 
macroscopic dynamics of the network of QIF neurons and postpone
the mathematical justification of its validity
to Sec.~\ref{validity}.

\begin{figure*}
\centerline{\includegraphics[width=160mm,clip=true]{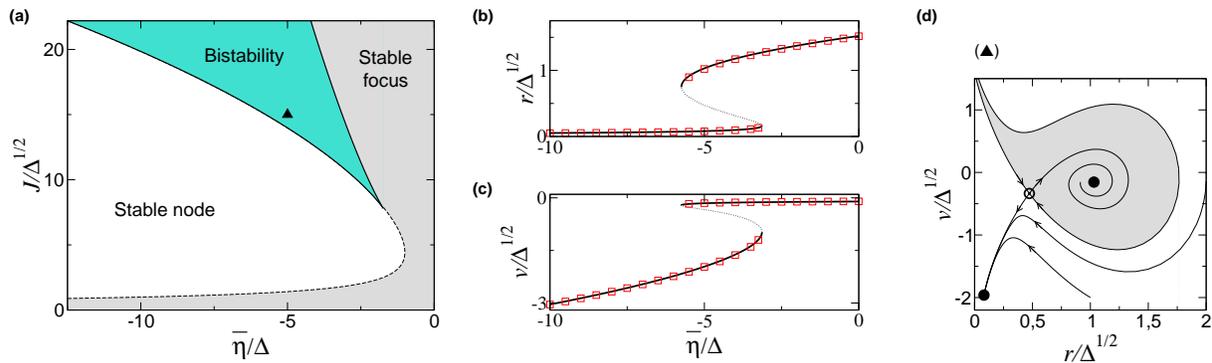}}
\caption{(color online). Analysis of the steady states of FREs \eqref{fre}. 
(a) Phase diagram: In the wedge-shaped cyan-shaded region 
there is bistability between a high and a low activity state. 
The boundary of the bistability region is the locus of a saddle-node 
bifurcation which is exactly obtained in parametric form: $(\bar \eta, J)_{SN}= 
[-\pi^2 r^2 - 3 \Delta^2 /(2\pi r)^2, 2\pi^2 r + \Delta^2/(2\pi^2 r^3)]$.
To the right of the dashed line, defined by 
$\bar\eta_f=-[J/(2\pi)]^2-(\pi\Delta /J)^2$, there is a stable focus
(shaded regions). (b) $r$-$\bar \eta$ and (c) $v$-$\bar \eta$ bifurcation 
diagrams for $J/\Delta^{1/2}=15$. Square symbols: Results obtained from numerical 
simulations of QIF neurons
(see Appendix A for details).
(d) Phase portrait of the system in 
the bistable region ($\bar\eta/\Delta=-5,J/\Delta^{1/2}=15$, triangle in panel (a)) with three fixed points: 
a stable focus (with its basin of attraction shaded), a stable node, 
and a saddle point.}
\label{Fig1}
\end{figure*}

\subsection{Macroscopic observables: Firing rate and mean membrane potential}
The half-width $x(\eta,t)$ of the LA has a particularly simple 
relation with the firing rate of the neuronal population 
(i.e.~the number of spikes per unit time). 
Indeed, the firing rate for each $\eta$ value at time $t$, $r(\eta,t)$,
can be computed by noting that neurons fire at a rate given 
by the probability flux at infinity: 
$r(\eta,t)= \rho(V\to \infty |\eta,t) \dot V (V\to \infty|\eta,t)$.
The limit $V\to\infty$ on the right hand side of this equation can be evaluated 
within the LA, and gives the simple identity 
\begin{equation}
x(\eta,t)=  \pi r(\eta,t). 
\label{x}
\end{equation}
The (total) firing rate $r(t)$ is then  
\begin{equation}
r(t)=\frac{1}{\pi}\int_{-\infty}^{\infty}x(\eta,t)g(\eta)d\eta.
\label{r}
\end{equation}
Additionally, the quantity $y(\eta,t)$ is, for each value of $\eta$,
the mean of the membrane potential:
\begin{equation}
y(\eta,t)= \mathrm{P.V.} 
\int_{-\infty}^{\infty} \rho(V|\eta,t) V \, dV.
\label{y}
\end{equation}
Here we take the Cauchy principal value, 
defined as $\mathrm{P.V.} \int_{-\infty}^{\infty} f(x)dx=
\lim_{R \to \infty}\int_{-R}^{R} f(x)dx$, to avoid the otherwise 
ill-defined integral. The mean membrane potential is then
\begin{equation}
v(t)= \int_{-\infty}^{\infty}  y(\eta,t)g(\eta)  d\eta.
\label{v}
\end{equation}
%

\subsection{Firing-rate equations}
Substituting the LA \eqref{la} into the continuity equation 
\eqref{cont}, we find that, for each value of $\eta$,
variables $x$ and $y$ must
obey two coupled equations which can be written in complex form as 
\begin{equation}
\partial_t w(\eta,t)= i\left[\eta + J s(t) -w(\eta,t)^2+I(t)\right]. 
\label{w}
\end{equation}
where $w(\eta,t)\equiv x(\eta,t) +i y(\eta,t)$.
Closing this equation requires expressing the mean synaptic activation
$s(t)$ as a function of $w(\eta,t)$. The simplest choice is to take the limit 
of infinitely fast synapses, $\tau\to 0$ in \eqref{s}, so that we get
an equality with the firing rate: $s(t)=r(t)$.
This allows for the system of QIF neurons 
\eqref{qif}-\eqref{s} to be exactly described by Eqs.~\eqref{w} and \eqref{r};
an infinite set of integro-differential equations if $g(\eta)$ is a 
continuous distribution.

Equation \eqref{w} is useful for general distributions $g(\eta)$ 
(see Appendix B), but a particularly sharp reduction in dimensionality
is achieved if $\eta$ is distributed according to 
a Lorentzian distribution of half-width $\Delta$ centered at $\bar \eta$: 
\begin{equation}
g(\eta)= \frac{1}{\pi} \frac{\Delta}{(\eta-\bar\eta)^2 +\Delta^2} .
\label{lorentzian}
\end{equation}
Note that this distribution accounts for the quenched variability in the 
external inputs. The fact that it is Lorentzian is unrelated to the LA 
for the density of membrane potentials.
Assuming \eqref{lorentzian} the integrals in \eqref{r} and \eqref{v} can be evaluated  
closing the integral contour in the complex $\eta$-plane 
and using the residue theorem\footnote{We make an analytic continuation of
$w(\eta,t)$ from real $\eta$ into complex-valued $\eta=\eta_r+i\eta_i$.
This is possible into the lower half-plane $\eta_i<0$,
since this guarantees the half-width $x(\eta,t)$ remains positive
zero: $\partial_t x(\eta,t)=- \eta_i >0$ at $x=0$.
Therefore we closed the integrals in \eqref{r} and \eqref{v} with 
an arc $|\eta| e^{i\vartheta}$ with
$|\eta|\to\infty$ and $\vartheta\in(-\pi,0)$. 
This contour encloses one pole of the 
Lorentzian distribution \eqref{lorentzian}, which written in partial
fractions reads: 
$g(\eta)=(2\pi i)^{-1}[(\eta-\bar\eta-i\Delta)^{-1}-
(\eta-\bar\eta+i\Delta)^{-1}]$.}.
Notably, the firing rate and the mean membrane potential  
depend only on the value of $w$ at 
the pole of $g(\eta)$ in the lower half $\eta$-plane:
$$
\pi r(t)+i v(t) = w(\bar\eta-i\Delta,t).
$$
As a result, we only need to evaluate \eqref{w} at $\eta=\bar\eta-i\Delta$, 
and thereby obtain a system of FREs composed of two ordinary differential equations
\begin{subequations}
\label{fre}
\begin{eqnarray}
\dot r &=& \Delta/\pi + 2  r v, \label{frea}\\ 
\dot v &=&   v^2 +   \bar \eta + J r + I(t) - \pi^2 r^2  . \label{freb}
\end{eqnarray}
\end{subequations}
This nonlinear system 
describes the macroscopic dynamics of the population of QIF neurons in 
terms of the population firing rate $r$ and mean membrane potential $v$. 

It is enlightening to compare mean-field model \eqref{fre} 
with the equations of the spiking neurons. 
Note that Eq.~\eqref{freb} resembles equations~\eqref{qif} and \eqref{qif2}
for the individual QIF neuron, but \emph{without spike resetting}. 
Indeed, the macroscopic firing-rate variable $r$ enters as a negative feedback term
in Eq.~\eqref{freb}, and impedes the explosive growth 
of the mean membrane potential $v$.

This negative feedback, combined with the 
coupling term on the right hand side of \eqref{frea}, describe the 
effective interaction between the firing rate and mean membrane potential 
at the network level. 
Therefore, the FREs \eqref{fre} 
describe the effect of the single-cell spike generation and reset mechanism 
at the network level.

In the following we examine the dynamics described by Eq.~\eqref{fre}, 
and show that they fully reproduce the macroscopic dynamics of 
the network of QIF neurons, even during episodes of strong 
spike synchrony.

\begin{figure*}
\begin{center}
\centerline{\includegraphics[width=165mm,clip=true]{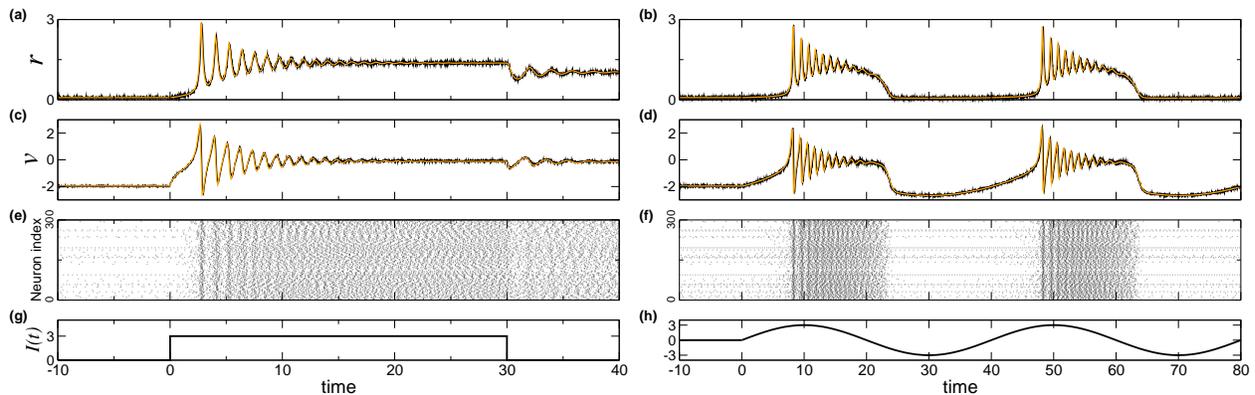}}
\caption{(color online). The transient dynamics of an ensemble of QIF model neurons 
\eqref{qif}-\eqref{qif2} are exactly described by the FREs \eqref{fre}. 
The instantaneous firing rate (Panels a,b) and mean membrane potential 
(Panels c,d) of the QIF neurons and the FREs are depicted in black and orange, respectively.
Panels (e,f) show the raster plots of 300 randomly selected QIF neurons
of the $N=10^4$  in the ensemble.
Left Panels (a,c,e,g): At time $t=0$, a 
current $I=3$ is applied to all neurons, and set to zero again at $t=30$; 
stimulus $I(t)$ shown in panel (g). 
Right panels (b,d,f,h): At time $t=0$ a sinusoidal current is applied to all neurons 
$I(t)=I_0\sin(\omega t)$, with $I_0=3$, $\omega=\pi/20$; stimulus $I(t)$ shown in panel (h). 
Parameters: $J=15$, $\bar\eta=-5, \Delta=1$.}
\label{Fig2}
\end{center}
\end{figure*}

\subsection{Analysis of the firing-rate equations} 

To begin with the analysis of Eq.~\eqref{fre}, we
first note that 
in the absence of forcing, $I(t)=0$, the only attractors of \eqref{fre} 
are fixed points. 
Figure~\ref{Fig1}(a) shows a phase 
diagram of the system as a function of the mean external drive $\bar{\eta}$ 
and synaptic weight $J$, both normalized by the width of the input 
distribution \footnote{The number of effective parameters can be reduced by 
nondimensionalizing the system as: $\tilde \eta= \bar \eta/\Delta$,
$\tilde J= J/\sqrt{\Delta}$, $(\tilde r, \tilde v)=(r,v)/ \sqrt{\Delta}$,
$\tilde t= \sqrt\Delta t$, and $\tilde I(\tilde t)=I(t/\sqrt\Delta)/\Delta$.}.  
There are three qualitatively distinct regions of the 
phase diagram: 1 - A single stable node 
corresponding to a low-activity state (white), 
2 - A single stable focus (spiral) 
generally corresponding to a high-activity state (gray), 
and 3 - A region of bistability between low and high firing rate (cyan; 
see a phase portrait of this region in Fig.~\ref{Fig1}(d)).  
Comparison of a sample bifurcation diagram of the fixed points
from numerical simulation of 
networks of QIF neurons with that obtained from the FREs \eqref{fre} 
shows an excellent correspondence, see Fig.~\ref{Fig1}(b,c). 

A similar phase diagram can be readily 
reproduced by traditional heuristic firing-rate models, 
with one significant qualitative difference: the presence of a 
stable focus ---and hence damped oscillations. 
Specifically, in the gray region of the 
phase diagram in Fig.~\ref{Fig1}(a), the system
undergoes oscillatory decay to the stable fixed point.
This oscillatory decay occurs as well for the high-activity state
over a large extent of the region of bistability (cyan), see e.g.~Fig.~\ref{Fig1}(d).

The presence of damped oscillations at the 
macroscopic level reflects the transitory synchronous firing of a 
fraction of the neurons in the ensemble. This behavior 
is common in spiking neuron models with 
weak noise, and is not captured by traditional firing rate 
models (see e.g. \cite{SOA13}).

\subsection{Analysis of the firing-rate equations: Non-stationary inputs}

To show that the FREs~\eqref{fre} fully describe the macroscopic 
response of the population of QIF neurons to time-varying stimuli 
(up to finite-size effects), 
we consider two types of stimulus $I(t)$: 1 - a step 
function and 2 - a sinusoidal forcing.  
In both cases we simulate the full system of 
QIF neurons and the FREs \eqref{fre}.
 
Figure~\ref{Fig2} shows the system's response to the two different inputs.  
In both cases the system is initially ($t<0$) in a bistable regime and 
set in the low activity state, with parameters corresponding to those of
Fig.~\ref{Fig1}(d).  Left panels of Fig.~\ref{Fig2} show the response
of the system to a  step current, applied at $t=0$. 
The applied current is such that the  system abandons the bistable region 
---see Fig.~\ref{Fig1}(a)--- and approaches the high activity state, which is
a stable focus. This is  clearly reflected in the time series
$r(t),~v(t)$, where the rate equations~\eqref{fre}  exactly predict 
the damped oscillations exhibited by the mean-field of the QIF neurons.
The raster plot in panel (e) shows the presence of the 
oscillations, which is due to the transitory synchronous firing of a 
large fraction of neurons in the population. 
Finally, at $t=30$, the current is removed and the system 
converges ---again, showing damped oscillations--- to the 
new location of the (focus) fixed point, which clearly  coexists with the 
stable node where it was originally placed ($t<0$).  

The right hand panels of Fig.~\ref{Fig2} show the response of the model to a
periodic current, which  drives the system
from one side of the bistable region to the other.  As a result, we
observe periodic bursting behavior when the system visits the stable
focus region of the phase diagram Fig.~\ref{Fig1}(a). 

To further illustrate the potential of the FREs~\eqref{fre} 
to predict and investigate complex dynamics in ensembles of spiking 
neurons, we present here a simple situation where the 
system of QIF neurons exhibits macroscopic chaos.
This is observed by increasing
the frequency $\omega$ of the sinusoidal driving, so that the system cannot 
trivially follow the stable fixed point at each cycle of the applied current.

\begin{figure}
\centerline{\includegraphics[width=70mm,clip=true]{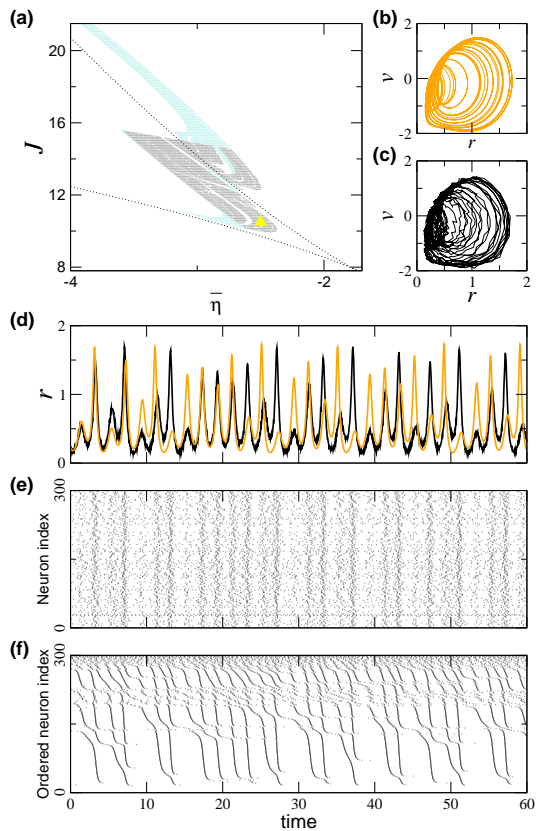}}
\caption{(color online). Firing rate model predicts the existence of macroscopic chaos
in a ensemble QIF neurons \eqref{qif} with an injected periodic current 
$I(t)=I_0 \sin(\omega t)$. (a) Phase diagram with the regions where chaos is found,
obtained using the rate model~\eqref{fre}. In the black-shaded 
region, there is only one chaotic attractor, whereas in the cyan region   
the chaotic attractor coexists with a periodic attractor. 
Dotted lines correspond to the bistability boundary of Fig.~\ref{Fig1}(a), 
depicted to facilitate comparison. 
(b) Chaotic trajectory obtained simulating the FREs \eqref{fre}; the Lyapunov
exponent is $\lambda=0.183\ldots$.
(c) Chaotic trajectory obtained from the QIF neurons 
---parameters corresponding to the (yellow) triangle symbol in 
panel (a). (d) Time series for the 
rate model (orange) and QIF model (black). 
(e) Raster plot corresponding to 300 randomly selected neurons. 
(f) Same raster plot as in (e), ordering neurons 
according to their intrinsic current $\eta_k$.   
Parameters: $I_0=3$, $\omega=\pi$; and
$\bar\eta=-2.5$, $J=10.5$ (triangle symbol in (a)).}
\label{Fig3}
\end{figure}

Figure 3(a) shows a phase diagram obtained using 
Eq.~\eqref{fre}. The shaded regions indicate
parameter values where the rate model has either a single chaotic attractor 
(in black) or a chaotic attractor coexisting with a
periodic orbit (in cyan). 
A trajectory on the chaotic attractor is depicted in Fig.~3(b), 
and the clearly irregular time series of the firing rate is 
shown in orange in Fig.~3(d).
Using the same parameters we performed numerical simulations of the
QIF neurons~\eqref{qif}-\eqref{qif2}, finding a similar attractor and irregular 
dynamics as in the rate model, see Figs.~3(c) and (d). To obtain the 
time series shown in Figs.~3(d), we ran the QIF neurons numerically 
and, after a long transient, 
at time $t=0$, the rate model \eqref{fre} was initiated with the values of $r$ and $v$
obtained from the population of QIF neurons. 
The time series of the two systems, which are initially 
close, rapidly diverge reflecting the chaotic nature of the system. 

Finally, to illustrate this chaotic state at the microscopic level, 
Fig.~3(e) shows the raster plot
for 300 randomly chosen neurons, corresponding to 
the time series in Fig.~3(d). 
The irregular firing of neurons in Fig.~3(e)
has some underlying structure that 
may be visualized ordering the same set of neurons according to 
their intrinsic currents as
$\eta_k<\eta_{k+1}$, with $k\in[1,300]$, see Fig.~3(f). 
Clearly, the maxima of the firing rate coincide with the synchronous 
firings of clusters of neurons with similar $\eta$ values. 
The size of these clusters is highly 
irregular in time, in concomitance with the chaotic behavior.

\subsection{Validity of the Lorentzian ansatz}
\label{validity}
Thus far, we have shown that the LA \eqref{la} solves the continuity 
equation \eqref{cont}, and confirmed that these solutions
agree with the numerical simulations of the original system of QIF 
neurons \eqref{qif}-\eqref{qif2}. Here we further clarify
why the LA holds for ensembles of QIF neurons.

Transforming the voltage of the QIF neuron into a phase via 
$V_j=\tan(\theta_j/2)$, the system \eqref{qif}-\eqref{qif2} 
becomes an ensemble of `theta neurons' \cite{EK86}:
\begin{equation}
\dot\theta_j= (1-\cos\theta_j)+ (1+\cos\theta_j)[\eta_j+ J s(t)+I(t)].
\label{ThN} 
\end{equation}
In the new phase variable $\theta\in[0,2\pi)$, the LA \eqref{la} becomes:
\begin{equation}
\tilde\rho(\theta \vert \eta, t)=\frac{1}{2\pi} \mbox{Re}
\left[ \frac{1+\alpha(\eta,t)e^{i\theta}}{1-\alpha(\eta,t) e^{i\theta}} \right],
\label{oa}
\end{equation}
where the function $\alpha(\eta,t)$ is related with  $w(\eta,t)$ as 
\begin{equation}
\alpha(\eta,t)=\frac{1-w(\eta,t)}{1+w(\eta,t)}.
\label{alpha} 
\end{equation}
Equations~\eqref{la} and \eqref{oa} are 
two representations of the so-called Poisson kernel on the half-plane 
and on the unit disk, respectively. These representations are 
related via equation \eqref{alpha}, that  
establishes a conformal mapping from the
half-plane $\mbox{Re}(w)\ge0$ onto the unit disk $|\alpha|\le1$. 
In the next section we show that variables $r$ and $v$ can be related, 
via the same conformal map, with the Kuramoto order parameter, 
which is a macroscopic measure of phase coherence \cite{Kur84,PRK01}.
 
The key observation supporting the applicability of the LA 
is the fact that equation~\eqref{oa} turns out to be the ansatz discovered 
in 2008 by Ott and Antonsen \cite{OA08}.
According to the Ott-Antonsen (OA) theory, in the thermodynamic 
limit the dynamics of the class of systems 
\begin{equation}
\partial_t\theta(\eta,t)=\Omega(\eta,t)+ \mathrm{Im}\left[H(\eta,t) 
e^{-i\theta}\right],
\label{wbh}
\end{equation}
generally converges to the OA manifold \eqref{oa}. 
In our case, for equation~\eqref{ThN}, 
we have $\Omega(\eta,t)=1+ \eta+ J s+ I$ and $H(\eta,t)=i(-1+\eta+Js+I)$. 
Thus far, the convergence of \eqref{wbh} to the OA manifold 
has been proven only for $H(\eta,t)=H(t)$. This includes the well-known 
Kuramoto and Winfree models \cite{OA09,*OHA11,PM14}, but not system 
\eqref{ThN}. However, there are theoretical arguments \cite{PR11} 
that strongly 
suggest that the OA manifold is also attracting for $\eta$-dependent 
$H$, as numerically confirmed by a number of recent papers 
using theta neurons \cite{LBS13,SLB14,Lai14} and other phase-oscillator 
models \cite{MP11,PM11,MP11p,IPM+13,IMS14}.
%

\subsection{Firing rate and Kuramoto order parameter}

There exists a mapping between the 
macroscopic variables $r$ and $v$ and the so-called 
Kuramoto order parameter $Z$.  
Equation~\eqref{alpha} relates, for each value of $\eta$,
the firing rate and the mean membrane potential, both contained in $w$, 
with the uniformity of the phase density, 
measured by  $\alpha$ ---note that $\alpha=0$ in \eqref{oa}
yields a perfectly uniform density. 
The Kuramoto order parameter is obtained 
by integrating $\alpha$ over the whole population as
\begin{eqnarray}
Z(t) & = & \int_{-\infty}^{\infty} g(\eta)\int_{0}^{2\pi} 
\tilde\rho(\theta|\eta,t) e^{i\theta}
d\theta \, d\eta \nonumber \\ 
&=&\int_{-\infty}^{\infty} g(\eta) \alpha^*(\eta,t) \,d\eta ,
\label{kop}
\end{eqnarray}
where we assumed that the density $\tilde\rho$ is the OA ansatz Eq.~\eqref{oa}.
The quantity $Z$ can be seen as the center of mass of a 
population of phases distributed across the unit circle $e^{i\theta}$. 

For a Lorentzian distribution of currents, $g(\eta)$, we may derive
the exact formula (see Appendix B)  
\begin{equation}
Z= \frac{1-W^*}{1+W^*}.
\label{W}
\end{equation}
relating the Kuramoto order parameter $Z$ with the firing-rate 
quantity $W\equiv\pi r + i v $. 
Figure 4 illustrates this conformal mapping of the 
right half-plane ($r\ge0$) onto  the unit disk ($|Z|\le1$).

\begin{figure}
\centerline{\includegraphics[width=75mm,clip=true]{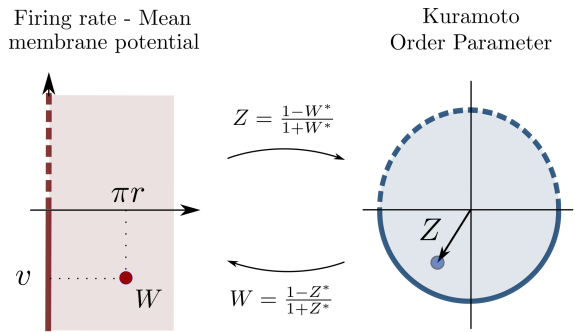}}
\caption{(color online). The conformal map \eqref{W} transforms the right half-plane
onto the unit disk. This transformation and its inverse 
define a one-to-one mapping between the Kuramoto order 
parameter $Z$ and the macroscopic quantities of the population of QIF neurons: 
$r$: firing rate, and $v$: mean membrane potential. 
Note that in the limit $Z=e^{i\Psi}$ 
(full coherence, $\tilde\rho(\theta)=\delta(\theta-\Psi)$) we
recover $v=i(1-e^{i\Psi})/(1+e^{i\Psi})=\tan(\Psi/2)$,
as the original mapping $V=\tan(\theta/2)$ between QIF- 
and theta-neurons dictates.}
\end{figure}

\section{Conclusions}

We have presented a method for deriving
firing rate equations for a network of heterogeneous QIF neurons, 
which is exact in the thermodynamic limit.
To our knowledge, the resulting system of ordinary differential
equations~\eqref{fre} represents the first exact
FREs of a network of spiking neurons.  

We would like to  emphasize that the derivation of the FREs does not 
rely on assumptions of weak coupling, separation of time-scales, 
averaging, or any other approximation. 
Rather, the assumptions underlying the
validity of equation~\eqref{fre} are that $i)$ the QIF 
neurons be all-to-all
connected, $ii)$ heterogeneity be quenched, and $iii)$ inputs 
be distributed according to a Lorentzian distribution.

The last two assumptions may seem particularly restrictive. 
Concerning $(iii)$, it must be stressed that for
\emph{arbitrary} distributions of quenched heterogeneity
the LA \eqref{la} remains generally valid, and therefore equation \eqref{w} 
can be used to discern the stability of the network states (see Appendix C). 
Furthermore, in Appendix C we also show numerical simulations 
using uniform- and Gaussian-distributed 
inputs, that reveal very similar macroscopic dynamics in response to time 
varying inputs. Even relaxing assumption $(ii)$, numerical simulations
with identical QIF neurons driven by external independent Gaussian noise 
sources show qualitative agreement with the FREs \eqref{fre} (see Appendix D).
In sum, the choice of a quenched Lorentzian distribution $g(\eta)$
is thus a mere mathematical convenience, whereas the
insights  gained from the resulting firing-rate description are valid
more generally. 

Our derivation represents a sharp departure from, and
a major advance compared to previous studies in several regards. 
Firstly, in the past it has only been possible to calculate the 
approximate firing rate of networks of spiking neurons 
for stationary states, or for weak deviations of such states 
\cite{Ger95,FM98,AB97i,Bru00,SHS03,OB11,SOA13}.   
The equations which result from these derivations are
difficult to solve, and typically require special numerical
methods. In contrast, FREs \eqref{fre} can be easily analyzed 
and simulated, and exactly reproduce the  behavior of the spiking 
network far from any fixed point and for arbitrary external currents. 
Secondly, firing-rate descriptions traditionally 
assume that the activity of a population of neurons
is equivalent to a set of uncorrelated stochastic processes with a
given rate, see e.g.~\cite{DA01,GK02}. 
However, in simulations of spiking networks it is well known that the response of the 
network to non-stationary inputs generally involves some degree of spike 
synchrony.  Recent theoretical work has sought to improve on classical 
rate models, which act as a low-pass filter of inputs, by fitting them with 
linear filters extracted from the corresponding Fokker-Planck equation for 
the network \cite{OB11,SOA13}.  These filters tend to generate damped oscillations 
when the external noise level is not too high, reflecting the presence of 
spike synchrony \cite{SHS03,SOA13}. The FREs \eqref{fre}  
also capture this phenomenon, and reveal that the underlying mechanism 
is an interplay between firing rate and subthreshold voltage.  
Furthermore, because these equations are 
exact, we can explore the full nonlinear response of the network,
such as  the generation of chaotic states of synchronous 
bursting as shown in Fig.~3.

Recently it has been shown that, not only Kuramoto-like models, but a much wider 
class of phase-models, evolve in the OA manifold defined by 
\eqref{oa}. Specifically, networks of pulse-coupled oscillators \cite{PM14} 
and theta-neurons \cite{LBS13,SLB14,Lai14} allow for an exact, low-dimensional description 
in terms of the Kuramoto order parameter \eqref{kop}. Although these later 
works use a finite-width phase coupling function that differs from the 
synaptic coupling~\eqref{s}, the obtained low-dimensional description in terms of 
the Kuramoto order parameter is analogous to ours, 
but in a different space \footnote{However, note that the phase diagrams 
in Luke et al. \cite{LBS13} do not fully agree with ours, and show additional 
instabilities. These bifurcations are associated with the finite width of 
the pulses, as it accurs in the Winfree model \cite{PM14}.}. 
Indeed, we showed that the OA ansatz  \eqref{oa} is related, 
via the nonlinear transformation of variables \eqref{alpha}, 
to the LA ansatz \eqref{la}. Remarkably, this transformation 
establishes an exact correspondence between the Kuramoto order parameter and 
a novel, biophysically maningful macroscopic 
observable which describes the firing rate and mean membrane potential of the 
neuronal network. Interestingly, the low-dimensional description in terms of 
firing rates  seems to be a more natural description for networks 
of spiking neurons, compared to that in terms 
of the Kuramoto order parameter. The firing rate equations \eqref{fre} take a 
surprisingly simple form in the LA manifold, what makes them a valuable 
tool to explore and and understand the mechanisms governing the macroscopic 
dynamics of neuronal networks.   

Finally, since the OA ansatz is the asymptotic solution for systems of the form in
equation~\eqref{wbh}, applying the change of variables
$V=\tan(\theta/2)$ automatically implies that the LA should hold for
populations governed by:
\begin{equation}
\dot V_j= A(\eta_j,t)+B(\eta_j,t) V_j + C(\eta_j,t)V_j^2,
\label{general_QIF}
\end{equation}
where $A$, $B$, and $C$ are related with $\Omega$ and $H$ as 
$A=[\Omega+\mathrm{Im}(H)]/2$, $B=-\mathrm{Re}(H)$,
$C=[\Omega-\mathrm{Im}(H)]/2$. Notably, equation~\eqref{general_QIF}
defines a wide family of ensembles of 
QIF neurons.

Therefore, the LA is actually valid for more general 
networks beyond the one investigated here 
---$\eta_j$ in Eq.~\eqref{general_QIF} may also be a vector 
containing several forms of disorder.
As particularly relevant cases, in Appendix E 
we provide the FREs governing the dynamics of an 
excitatory network with both, heterogeneous inputs $\eta$ \emph{and} 
synaptic weights $J$, as well as pair of 
interacting excitatory and inhibitory populations of 
QIF neurons. 
In addition, according to Eq.~\eqref{general_QIF}, 
the LA is also valid if synapses are  modeled as conductances, 
in which case the reversal potentials may be distributed as well. 
Moreover, the role of gap-junctions 
or synaptic kinetics 
may be considered within the same framework.


\begin{thebibliography}{69}%
\makeatletter
\providecommand \@ifxundefined [1]{%
 \@ifx{#1\undefined}
}%
\providecommand \@ifnum [1]{%
 \ifnum #1\expandafter \@firstoftwo
 \else \expandafter \@secondoftwo
 \fi
}%
\providecommand \@ifx [1]{%
 \ifx #1\expandafter \@firstoftwo
 \else \expandafter \@secondoftwo
 \fi
}%
\providecommand \natexlab [1]{#1}%
\providecommand \enquote  [1]{``#1''}%
\providecommand \bibnamefont  [1]{#1}%
\providecommand \bibfnamefont [1]{#1}%
\providecommand \citenamefont [1]{#1}%
\providecommand \href@noop [0]{\@secondoftwo}%
\providecommand \href [0]{\begingroup \@sanitize@url \@href}%
\providecommand \@href[1]{\@@startlink{#1}\@@href}%
\providecommand \@@href[1]{\endgroup#1\@@endlink}%
\providecommand \@sanitize@url [0]{\catcode `\\12\catcode `\$12\catcode
  `\&12\catcode `\#12\catcode `\^12\catcode `\_12\catcode `\%12\relax}%
\providecommand \@@startlink[1]{}%
\providecommand \@@endlink[0]{}%
\providecommand \url  [0]{\begingroup\@sanitize@url \@url }%
\providecommand \@url [1]{\endgroup\@href {#1}{\urlprefix }}%
\providecommand \urlprefix  [0]{URL }%
\providecommand \Eprint [0]{\href }%
\providecommand \doibase [0]{http://dx.doi.org/}%
\providecommand \selectlanguage [0]{\@gobble}%
\providecommand \bibinfo  [0]{\@secondoftwo}%
\providecommand \bibfield  [0]{\@secondoftwo}%
\providecommand \translation [1]{[#1]}%
\providecommand \BibitemOpen [0]{}%
\providecommand \bibitemStop [0]{}%
\providecommand \bibitemNoStop [0]{.\EOS\space}%
\providecommand \EOS [0]{\spacefactor3000\relax}%
\providecommand \BibitemShut  [1]{\csname bibitem#1\endcsname}%
\let\auto@bib@innerbib\@empty
\bibitem [{\citenamefont {Averbeck}\ \emph {et~al.}(2006)\citenamefont
  {Averbeck}, \citenamefont {Latham},\ and\ \citenamefont {Pouget}}]{ALP06}%
  \BibitemOpen
  \bibfield  {author} {\bibinfo {author} {\bibfnamefont {B.~B.}\ \bibnamefont
  {Averbeck}}, \bibinfo {author} {\bibfnamefont {P.~E.}\ \bibnamefont
  {Latham}}, \ and\ \bibinfo {author} {\bibfnamefont {A.}~\bibnamefont
  {Pouget}},\ }\bibfield  {title} {\enquote {\bibinfo {title} {Neural
  correlations, population coding and computation},}\ }\href@noop {} {\bibfield
   {journal} {\bibinfo  {journal} {Nature Reviews Neuroscience}\ }\textbf
  {\bibinfo {volume} {7}},\ \bibinfo {pages} {358--366} (\bibinfo {year}
  {2006})}\BibitemShut {NoStop}%
\bibitem [{\citenamefont {Hodgkin}\ and\ \citenamefont {Huxley}(1952)}]{HH52}%
  \BibitemOpen
  \bibfield  {author} {\bibinfo {author} {\bibfnamefont {A.~L.}\ \bibnamefont
  {Hodgkin}}\ and\ \bibinfo {author} {\bibfnamefont {A.~F.}\ \bibnamefont
  {Huxley}},\ }\bibfield  {title} {\enquote {\bibinfo {title} {A quantitative
  description of membrane current and its application to conduction and
  excitation in nerve},}\ }\href@noop {} {\bibfield  {journal} {\bibinfo
  {journal} {J Physiol.}\ }\textbf {\bibinfo {volume} {117}},\ \bibinfo {pages}
  {500--544} (\bibinfo {year} {1952})}\BibitemShut {NoStop}%
\bibitem [{\citenamefont {Brette}\ \emph {et~al.}(2007)\citenamefont {Brette},
  \citenamefont {Rudolph}, \citenamefont {Carnevale}, \citenamefont {Hines},
  \citenamefont {Beeman}, \citenamefont {Bower}, \citenamefont {Diesmann},
  \citenamefont {Morrison}, \citenamefont {Goodman}, \citenamefont {Harris~Jr}
  \emph {et~al.}}]{BRC+07}%
  \BibitemOpen
  \bibfield  {author} {\bibinfo {author} {\bibfnamefont {Romain}\ \bibnamefont
  {Brette}}, \bibinfo {author} {\bibfnamefont {Michelle}\ \bibnamefont
  {Rudolph}}, \bibinfo {author} {\bibfnamefont {Ted}\ \bibnamefont
  {Carnevale}}, \bibinfo {author} {\bibfnamefont {Michael}\ \bibnamefont
  {Hines}}, \bibinfo {author} {\bibfnamefont {David}\ \bibnamefont {Beeman}},
  \bibinfo {author} {\bibfnamefont {James~M}\ \bibnamefont {Bower}}, \bibinfo
  {author} {\bibfnamefont {Markus}\ \bibnamefont {Diesmann}}, \bibinfo {author}
  {\bibfnamefont {Abigail}\ \bibnamefont {Morrison}}, \bibinfo {author}
  {\bibfnamefont {Philip~H}\ \bibnamefont {Goodman}}, \bibinfo {author}
  {\bibfnamefont {Frederick~C}\ \bibnamefont {Harris~Jr}},  \emph {et~al.},\
  }\bibfield  {title} {\enquote {\bibinfo {title} {Simulation of networks of
  spiking neurons: a review of tools and strategies},}\ }\href@noop {}
  {\bibfield  {journal} {\bibinfo  {journal} {Journal of computational
  neuroscience}\ }\textbf {\bibinfo {volume} {23}},\ \bibinfo {pages}
  {349--398} (\bibinfo {year} {2007})}\BibitemShut {NoStop}%
\bibitem [{\citenamefont {Izhikevich}\ and\ \citenamefont
  {Edelman}(2008)}]{IE08}%
  \BibitemOpen
  \bibfield  {author} {\bibinfo {author} {\bibfnamefont {E.~M.}\ \bibnamefont
  {Izhikevich}}\ and\ \bibinfo {author} {\bibfnamefont {G.~M.}\ \bibnamefont
  {Edelman}},\ }\bibfield  {title} {\enquote {\bibinfo {title} {Large-scale
  model of mammalian thalamocortical systems},}\ }\href@noop {} {\bibfield
  {journal} {\bibinfo  {journal} {Proc. Nat. Acad. Sci.}\ }\textbf {\bibinfo
  {volume} {105}},\ \bibinfo {pages} {3593--3598} (\bibinfo {year}
  {2008})}\BibitemShut {NoStop}%
\bibitem [{\citenamefont {Izhikevich}(2007)}]{Izh07}%
  \BibitemOpen
  \bibfield  {author} {\bibinfo {author} {\bibfnamefont {E.~M.}\ \bibnamefont
  {Izhikevich}},\ }\href@noop {} {\emph {\bibinfo {title} {Dynamical Systems in
  Neuroscience.}}}\ (\bibinfo  {publisher} {The MIT Press},\ \bibinfo {address}
  {Cambridge, Massachusetts},\ \bibinfo {year} {2007})\BibitemShut {NoStop}%
\bibitem [{\citenamefont {Wilson}\ and\ \citenamefont {Cowan}(1972)}]{WC72}%
  \BibitemOpen
  \bibfield  {author} {\bibinfo {author} {\bibfnamefont {Hugh~R}\ \bibnamefont
  {Wilson}}\ and\ \bibinfo {author} {\bibfnamefont {Jack~D}\ \bibnamefont
  {Cowan}},\ }\bibfield  {title} {\enquote {\bibinfo {title} {Excitatory and
  inhibitory interactions in localized populations of model neurons},}\
  }\href@noop {} {\bibfield  {journal} {\bibinfo  {journal} {Biophys. J.}\
  }\textbf {\bibinfo {volume} {12}},\ \bibinfo {pages} {1--24} (\bibinfo {year}
  {1972})}\BibitemShut {NoStop}%
\bibitem [{\citenamefont {Amari}(1974)}]{Ama74}%
  \BibitemOpen
  \bibfield  {author} {\bibinfo {author} {\bibfnamefont {Shun-ichi}\
  \bibnamefont {Amari}},\ }\bibfield  {title} {\enquote {\bibinfo {title} {A
  method of statistical neurodynamics},}\ }\href@noop {} {\bibfield  {journal}
  {\bibinfo  {journal} {Kybernetik}\ }\textbf {\bibinfo {volume} {14}},\
  \bibinfo {pages} {201--215} (\bibinfo {year} {1974})}\BibitemShut {NoStop}%
\bibitem [{\citenamefont {Freeman}(1975)}]{Fre75}%
  \BibitemOpen
  \bibfield  {author} {\bibinfo {author} {\bibfnamefont {Walter~J}\
  \bibnamefont {Freeman}},\ }\href@noop {} {\emph {\bibinfo {title} {Mass
  action in the nervous system}}}\ (\bibinfo  {publisher} {Academic Press, New
  York},\ \bibinfo {year} {1975})\BibitemShut {NoStop}%
\bibitem [{\citenamefont {Gerstner}(1995)}]{Ger95}%
  \BibitemOpen
  \bibfield  {author} {\bibinfo {author} {\bibfnamefont {W.}~\bibnamefont
  {Gerstner}},\ }\bibfield  {title} {\enquote {\bibinfo {title} {Time structure
  of the activity in neural network models},}\ }\href {\doibase
  10.1103/PhysRevE.51.738} {\bibfield  {journal} {\bibinfo  {journal} {Phys.
  Rev. E}\ }\textbf {\bibinfo {volume} {51}},\ \bibinfo {pages} {738--758}
  (\bibinfo {year} {1995})}\BibitemShut {NoStop}%
\bibitem [{\citenamefont {Fusi}\ and\ \citenamefont {Mattia}(1998)}]{FM98}%
  \BibitemOpen
  \bibfield  {author} {\bibinfo {author} {\bibfnamefont {S.}~\bibnamefont
  {Fusi}}\ and\ \bibinfo {author} {\bibfnamefont {M.}~\bibnamefont {Mattia}},\
  }\bibfield  {title} {\enquote {\bibinfo {title} {Collective behavior of
  networks with linear (vlsi) integrate-and-fire neurons},}\ }\href@noop {}
  {\bibfield  {journal} {\bibinfo  {journal} {Neural Comput.}\ }\textbf
  {\bibinfo {volume} {11}},\ \bibinfo {pages} {633--652} (\bibinfo {year}
  {1998})}\BibitemShut {NoStop}%
\bibitem [{\citenamefont {Amit}\ and\ \citenamefont {Brunel}(1997)}]{AB97i}%
  \BibitemOpen
  \bibfield  {author} {\bibinfo {author} {\bibfnamefont {Daniel~J}\
  \bibnamefont {Amit}}\ and\ \bibinfo {author} {\bibfnamefont {Nicolas}\
  \bibnamefont {Brunel}},\ }\bibfield  {title} {\enquote {\bibinfo {title}
  {Model of global spontaneous activity and local structured activity during
  delay periods in the cerebral cortex.}}\ }\href@noop {} {\bibfield  {journal}
  {\bibinfo  {journal} {Cerebral Cortex}\ }\textbf {\bibinfo {volume} {7}},\
  \bibinfo {pages} {237--252} (\bibinfo {year} {1997})}\BibitemShut {NoStop}%
\bibitem [{\citenamefont {Brunel}(2000)}]{Bru00}%
  \BibitemOpen
  \bibfield  {author} {\bibinfo {author} {\bibfnamefont {N.}~\bibnamefont
  {Brunel}},\ }\bibfield  {title} {\enquote {\bibinfo {title} {Dynamics of
  sparsely connected networks of excitatory and inhibitory spiking neurons},}\
  }\href@noop {} {\bibfield  {journal} {\bibinfo  {journal} {Journal of
  computational neuroscience}\ }\textbf {\bibinfo {volume} {8}},\ \bibinfo
  {pages} {183--208} (\bibinfo {year} {2000})}\BibitemShut {NoStop}%
\bibitem [{\citenamefont {Dayan}\ and\ \citenamefont {Abbott}(2001)}]{DA01}%
  \BibitemOpen
  \bibfield  {author} {\bibinfo {author} {\bibfnamefont {P.}~\bibnamefont
  {Dayan}}\ and\ \bibinfo {author} {\bibfnamefont {L.~F.}\ \bibnamefont
  {Abbott}},\ }\href@noop {} {\emph {\bibinfo {title} {Theoretical
  neuroscience}}}\ (\bibinfo  {publisher} {Cambridge, MA: MIT Press},\ \bibinfo
  {year} {2001})\BibitemShut {NoStop}%
\bibitem [{\citenamefont {Gerstner}\ and\ \citenamefont
  {Kistler}(2002)}]{GK02}%
  \BibitemOpen
  \bibfield  {author} {\bibinfo {author} {\bibfnamefont {W.}~\bibnamefont
  {Gerstner}}\ and\ \bibinfo {author} {\bibfnamefont {W.~M.}\ \bibnamefont
  {Kistler}},\ }\href@noop {} {\emph {\bibinfo {title} {Spiking neuron models:
  Single neurons, populations, plasticity}}}\ (\bibinfo  {publisher} {Cambridge
  university press},\ \bibinfo {year} {2002})\BibitemShut {NoStop}%
\bibitem [{\citenamefont {Shriki}\ \emph {et~al.}(2003)\citenamefont {Shriki},
  \citenamefont {Hansel},\ and\ \citenamefont {Sompolinsky}}]{SHS03}%
  \BibitemOpen
  \bibfield  {author} {\bibinfo {author} {\bibfnamefont {Oren}\ \bibnamefont
  {Shriki}}, \bibinfo {author} {\bibfnamefont {David}\ \bibnamefont {Hansel}},
  \ and\ \bibinfo {author} {\bibfnamefont {Haim}\ \bibnamefont {Sompolinsky}},\
  }\bibfield  {title} {\enquote {\bibinfo {title} {Rate models for
  conductance-based cortical neuronal networks},}\ }\href@noop {} {\bibfield
  {journal} {\bibinfo  {journal} {Neural Comput.}\ }\textbf {\bibinfo {volume}
  {15}},\ \bibinfo {pages} {1809--1841} (\bibinfo {year} {2003})}\BibitemShut
  {NoStop}%
\bibitem [{\citenamefont {Coombes}(2005)}]{Coo05}%
  \BibitemOpen
  \bibfield  {author} {\bibinfo {author} {\bibfnamefont {Stephen}\ \bibnamefont
  {Coombes}},\ }\bibfield  {title} {\enquote {\bibinfo {title} {Waves, bumps,
  and patterns in neural field theories},}\ }\href@noop {} {\bibfield
  {journal} {\bibinfo  {journal} {Biol. Cybern.}\ }\textbf {\bibinfo {volume}
  {93}},\ \bibinfo {pages} {91--108} (\bibinfo {year} {2005})}\BibitemShut
  {NoStop}%
\bibitem [{\citenamefont {Roxin}\ \emph {et~al.}(2005)\citenamefont {Roxin},
  \citenamefont {Brunel},\ and\ \citenamefont {Hansel}}]{RBH05}%
  \BibitemOpen
  \bibfield  {author} {\bibinfo {author} {\bibfnamefont {A.}~\bibnamefont
  {Roxin}}, \bibinfo {author} {\bibfnamefont {N.}~\bibnamefont {Brunel}}, \
  and\ \bibinfo {author} {\bibfnamefont {D.}~\bibnamefont {Hansel}},\
  }\bibfield  {title} {\enquote {\bibinfo {title} {Role of delays in shaping
  spatiotemporal dynamics of neuronal activity in large networks},}\
  }\href@noop {} {\bibfield  {journal} {\bibinfo  {journal} {Phys. Rev. Lett.}\
  }\textbf {\bibinfo {volume} {94}},\ \bibinfo {pages} {238103} (\bibinfo
  {year} {2005})}\BibitemShut {NoStop}%
\bibitem [{\citenamefont {Deco}\ \emph {et~al.}(2008)\citenamefont {Deco},
  \citenamefont {Jirsa}, \citenamefont {Robinson}, \citenamefont {Breakspear},\
  and\ \citenamefont {Friston}}]{DJR+08}%
  \BibitemOpen
  \bibfield  {author} {\bibinfo {author} {\bibfnamefont {G.}~\bibnamefont
  {Deco}}, \bibinfo {author} {\bibfnamefont {V.~K.}\ \bibnamefont {Jirsa}},
  \bibinfo {author} {\bibfnamefont {P.~A.}\ \bibnamefont {Robinson}}, \bibinfo
  {author} {\bibfnamefont {M.}~\bibnamefont {Breakspear}}, \ and\ \bibinfo
  {author} {\bibfnamefont {K.}~\bibnamefont {Friston}},\ }\bibfield  {title}
  {\enquote {\bibinfo {title} {The dynamic brain: from spiking neurons to
  neural masses and cortical fields},}\ }\href@noop {} {\bibfield  {journal}
  {\bibinfo  {journal} {PLoS Computational biology}\ }\textbf {\bibinfo
  {volume} {4}},\ \bibinfo {pages} {e1000092} (\bibinfo {year}
  {2008})}\BibitemShut {NoStop}%
\bibitem [{\citenamefont {Ermentrout}\ and\ \citenamefont
  {Terman}(2010)}]{ET10}%
  \BibitemOpen
  \bibfield  {author} {\bibinfo {author} {\bibfnamefont {G.~B.}\ \bibnamefont
  {Ermentrout}}\ and\ \bibinfo {author} {\bibfnamefont {D.~H.}\ \bibnamefont
  {Terman}},\ }\href@noop {} {\emph {\bibinfo {title} {Mathematical foundations
  of neuroscience}}},\ Vol.~\bibinfo {volume} {64}\ (\bibinfo  {publisher}
  {Springer},\ \bibinfo {year} {2010})\BibitemShut {NoStop}%
\bibitem [{\citenamefont {Ostojic}\ and\ \citenamefont {Brunel}(2011)}]{OB11}%
  \BibitemOpen
  \bibfield  {author} {\bibinfo {author} {\bibfnamefont {S.}~\bibnamefont
  {Ostojic}}\ and\ \bibinfo {author} {\bibfnamefont {N.}~\bibnamefont
  {Brunel}},\ }\bibfield  {title} {\enquote {\bibinfo {title} {From spiking
  neuron models to linear-nonlinear models},}\ }\href@noop {} {\bibfield
  {journal} {\bibinfo  {journal} {PLoS Comput. Biol.}\ }\textbf {\bibinfo
  {volume} {7}},\ \bibinfo {pages} {e1001056} (\bibinfo {year}
  {2011})}\BibitemShut {NoStop}%
\bibitem [{\citenamefont {Roxin}\ and\ \citenamefont
  {Montbri{\'o}}(2011)}]{RM11}%
  \BibitemOpen
  \bibfield  {author} {\bibinfo {author} {\bibfnamefont {A.}~\bibnamefont
  {Roxin}}\ and\ \bibinfo {author} {\bibfnamefont {E.}~\bibnamefont
  {Montbri{\'o}}},\ }\bibfield  {title} {\enquote {\bibinfo {title} {How
  effective delays shape oscillatory dynamics in neuronal networks},}\
  }\href@noop {} {\bibfield  {journal} {\bibinfo  {journal} {Physica D}\
  }\textbf {\bibinfo {volume} {240}},\ \bibinfo {pages} {323--345} (\bibinfo
  {year} {2011})}\BibitemShut {NoStop}%
\bibitem [{\citenamefont {Schaffer}\ \emph {et~al.}(2013)\citenamefont
  {Schaffer}, \citenamefont {Ostojic},\ and\ \citenamefont {Abbott}}]{SOA13}%
  \BibitemOpen
  \bibfield  {author} {\bibinfo {author} {\bibfnamefont {Evan~S}\ \bibnamefont
  {Schaffer}}, \bibinfo {author} {\bibfnamefont {Srdjan}\ \bibnamefont
  {Ostojic}}, \ and\ \bibinfo {author} {\bibfnamefont {LF}~\bibnamefont
  {Abbott}},\ }\bibfield  {title} {\enquote {\bibinfo {title} {A complex-valued
  firing-rate model that approximates the dynamics of spiking networks},}\
  }\href@noop {} {\bibfield  {journal} {\bibinfo  {journal} {PLoS Comput.
  Biol.}\ }\textbf {\bibinfo {volume} {9}},\ \bibinfo {pages} {e1003301}
  (\bibinfo {year} {2013})}\BibitemShut {NoStop}%
\bibitem [{\citenamefont {Hopfield}(1984)}]{Hop84}%
  \BibitemOpen
  \bibfield  {author} {\bibinfo {author} {\bibfnamefont {J.~J.}\ \bibnamefont
  {Hopfield}},\ }\bibfield  {title} {\enquote {\bibinfo {title} {Neurons with
  graded response have collective computational properties like those of
  two-state neurons},}\ }\href@noop {} {\bibfield  {journal} {\bibinfo
  {journal} {Proceedings of the national academy of sciences}\ }\textbf
  {\bibinfo {volume} {81}},\ \bibinfo {pages} {3088--3092} (\bibinfo {year}
  {1984})}\BibitemShut {NoStop}%
\bibitem [{\citenamefont {Mongillo}\ \emph {et~al.}(2008)\citenamefont
  {Mongillo}, \citenamefont {Barak},\ and\ \citenamefont {Tsodyks}}]{MBT08}%
  \BibitemOpen
  \bibfield  {author} {\bibinfo {author} {\bibfnamefont {G.}~\bibnamefont
  {Mongillo}}, \bibinfo {author} {\bibfnamefont {O.}~\bibnamefont {Barak}}, \
  and\ \bibinfo {author} {\bibfnamefont {M.}~\bibnamefont {Tsodyks}},\
  }\bibfield  {title} {\enquote {\bibinfo {title} {Synaptic theory of working
  memory},}\ }\href@noop {} {\bibfield  {journal} {\bibinfo  {journal}
  {Science}\ }\textbf {\bibinfo {volume} {319}},\ \bibinfo {pages} {1543--1546}
  (\bibinfo {year} {2008})}\BibitemShut {NoStop}%
\bibitem [{\citenamefont {Ben-Yishai}\ \emph {et~al.}(1995)\citenamefont
  {Ben-Yishai}, \citenamefont {Bar-Or},\ and\ \citenamefont
  {Sompolinsky}}]{BLS95}%
  \BibitemOpen
  \bibfield  {author} {\bibinfo {author} {\bibfnamefont {R}~\bibnamefont
  {Ben-Yishai}}, \bibinfo {author} {\bibfnamefont {R~Lev}\ \bibnamefont
  {Bar-Or}}, \ and\ \bibinfo {author} {\bibfnamefont {H}~\bibnamefont
  {Sompolinsky}},\ }\bibfield  {title} {\enquote {\bibinfo {title} {Theory of
  orientation tuning in visual cortex.}}\ }\href@noop {} {\bibfield  {journal}
  {\bibinfo  {journal} {Proc. Nat. Acad. Sci.}\ }\textbf {\bibinfo {volume}
  {92}},\ \bibinfo {pages} {3844--3848} (\bibinfo {year} {1995})}\BibitemShut
  {NoStop}%
\bibitem [{\citenamefont {Hansel}\ and\ \citenamefont
  {Sompolinsky}(1998)}]{HS98}%
  \BibitemOpen
  \bibfield  {author} {\bibinfo {author} {\bibfnamefont {David}\ \bibnamefont
  {Hansel}}\ and\ \bibinfo {author} {\bibfnamefont {Haim}\ \bibnamefont
  {Sompolinsky}},\ }\bibfield  {title} {\enquote {\bibinfo {title} {Modeling
  feature selectivity in local cortical circuits},}\ }in\ \href@noop {} {\emph
  {\bibinfo {booktitle} {Methods in Neuronal Modelling: From Ions to
  Networks}}},\ \bibinfo {editor} {edited by\ \bibinfo {editor} {\bibfnamefont
  {C.}~\bibnamefont {Koch}}\ and\ \bibinfo {editor} {\bibfnamefont
  {I.}~\bibnamefont {Segev}}}\ (\bibinfo  {publisher} {MIT Press},\ \bibinfo
  {address} {Cambridge},\ \bibinfo {year} {1998})\ pp.\ \bibinfo {pages}
  {499--567}\BibitemShut {NoStop}%
\bibitem [{\citenamefont {Moreno-Bote}\ \emph {et~al.}(2007)\citenamefont
  {Moreno-Bote}, \citenamefont {Rinzel},\ and\ \citenamefont {Rubin}}]{MRR07}%
  \BibitemOpen
  \bibfield  {author} {\bibinfo {author} {\bibfnamefont {Rub{\'e}n}\
  \bibnamefont {Moreno-Bote}}, \bibinfo {author} {\bibfnamefont {John}\
  \bibnamefont {Rinzel}}, \ and\ \bibinfo {author} {\bibfnamefont {Nava}\
  \bibnamefont {Rubin}},\ }\bibfield  {title} {\enquote {\bibinfo {title}
  {Noise-induced alternations in an attractor network model of perceptual
  bistability},}\ }\href@noop {} {\bibfield  {journal} {\bibinfo  {journal} {J.
  Neurophysiol.}\ }\textbf {\bibinfo {volume} {98}},\ \bibinfo {pages}
  {1125--1139} (\bibinfo {year} {2007})}\BibitemShut {NoStop}%
\bibitem [{\citenamefont {Zhang}(1996)}]{Zha96}%
  \BibitemOpen
  \bibfield  {author} {\bibinfo {author} {\bibfnamefont {K.}~\bibnamefont
  {Zhang}},\ }\bibfield  {title} {\enquote {\bibinfo {title} {Representation of
  spatial orientation by the intrinsic dynamics of the head-direction cell
  ensemble: a theory},}\ }\href@noop {} {\bibfield  {journal} {\bibinfo
  {journal} {J. Neurosci.}\ }\textbf {\bibinfo {volume} {16}},\ \bibinfo
  {pages} {2112--2126} (\bibinfo {year} {1996})}\BibitemShut {NoStop}%
\bibitem [{\citenamefont {Wong}\ and\ \citenamefont {Wang}(2006)}]{WW06}%
  \BibitemOpen
  \bibfield  {author} {\bibinfo {author} {\bibfnamefont {K.-F.}\ \bibnamefont
  {Wong}}\ and\ \bibinfo {author} {\bibfnamefont {X.-J.}\ \bibnamefont
  {Wang}},\ }\bibfield  {title} {\enquote {\bibinfo {title} {A recurrent
  network mechanism of time integration in perceptual decisions},}\ }\href@noop
  {} {\bibfield  {journal} {\bibinfo  {journal} {J. Neurosci.}\ }\textbf
  {\bibinfo {volume} {26}},\ \bibinfo {pages} {1314--1328} (\bibinfo {year}
  {2006})}\BibitemShut {NoStop}%
\bibitem [{\citenamefont {Abeles}(1991)}]{Abe91}%
  \BibitemOpen
  \bibfield  {author} {\bibinfo {author} {\bibfnamefont {Moshe}\ \bibnamefont
  {Abeles}},\ }\href@noop {} {\emph {\bibinfo {title} {Corticonics: Neural
  circuits of the cerebral cortex}}}\ (\bibinfo  {publisher} {Cambridge
  University Press},\ \bibinfo {year} {1991})\BibitemShut {NoStop}%
\bibitem [{\citenamefont {Varela}\ \emph {et~al.}(2001)\citenamefont {Varela},
  \citenamefont {Lachaux}, \citenamefont {Rodriguez},\ and\ \citenamefont
  {Martinerie}}]{VLR01}%
  \BibitemOpen
  \bibfield  {author} {\bibinfo {author} {\bibfnamefont {F.}~\bibnamefont
  {Varela}}, \bibinfo {author} {\bibfnamefont {J.-P.}\ \bibnamefont {Lachaux}},
  \bibinfo {author} {\bibfnamefont {E.}~\bibnamefont {Rodriguez}}, \ and\
  \bibinfo {author} {\bibfnamefont {J.}~\bibnamefont {Martinerie}},\ }\bibfield
   {title} {\enquote {\bibinfo {title} {The brainweb: phase synchronization and
  large-scale integration},}\ }\href@noop {} {\bibfield  {journal} {\bibinfo
  {journal} {Nature reviews neuroscience}\ }\textbf {\bibinfo {volume} {2}},\
  \bibinfo {pages} {229--239} (\bibinfo {year} {2001})}\BibitemShut {NoStop}%
\bibitem [{\citenamefont {Engel}\ and\ \citenamefont {Singer}(2001)}]{ES01}%
  \BibitemOpen
  \bibfield  {author} {\bibinfo {author} {\bibfnamefont {Andreas~K}\
  \bibnamefont {Engel}}\ and\ \bibinfo {author} {\bibfnamefont {Wolf}\
  \bibnamefont {Singer}},\ }\bibfield  {title} {\enquote {\bibinfo {title}
  {Temporal binding and the neural correlates of sensory awareness},}\
  }\href@noop {} {\bibfield  {journal} {\bibinfo  {journal} {Trends in
  cognitive sciences}\ }\textbf {\bibinfo {volume} {5}},\ \bibinfo {pages}
  {16--25} (\bibinfo {year} {2001})}\BibitemShut {NoStop}%
\bibitem [{\citenamefont {Ward}(2003)}]{War03}%
  \BibitemOpen
  \bibfield  {author} {\bibinfo {author} {\bibfnamefont {L.~M.}\ \bibnamefont
  {Ward}},\ }\bibfield  {title} {\enquote {\bibinfo {title} {Synchronous neural
  oscillations and cognitive processes},}\ }\href@noop {} {\bibfield  {journal}
  {\bibinfo  {journal} {Trends in cognitive sciences}\ }\textbf {\bibinfo
  {volume} {7}},\ \bibinfo {pages} {553--559} (\bibinfo {year}
  {2003})}\BibitemShut {NoStop}%
\bibitem [{\citenamefont {Fries}(2005)}]{Fri05}%
  \BibitemOpen
  \bibfield  {author} {\bibinfo {author} {\bibfnamefont {P.}~\bibnamefont
  {Fries}},\ }\bibfield  {title} {\enquote {\bibinfo {title} {A mechanism for
  cognitive dynamics: neuronal communication through neuronal coherence},}\
  }\href@noop {} {\bibfield  {journal} {\bibinfo  {journal} {Trends in
  cognitive sciences}\ }\textbf {\bibinfo {volume} {9}},\ \bibinfo {pages}
  {474--480} (\bibinfo {year} {2005})}\BibitemShut {NoStop}%
\bibitem [{\citenamefont {Buzsaki}(2006)}]{Buz06}%
  \BibitemOpen
  \bibfield  {author} {\bibinfo {author} {\bibfnamefont {Gyorgy}\ \bibnamefont
  {Buzsaki}},\ }\href@noop {} {\emph {\bibinfo {title} {Rhythms of the
  Brain}}}\ (\bibinfo  {publisher} {Oxford University Press},\ \bibinfo {year}
  {2006})\BibitemShut {NoStop}%
\bibitem [{\citenamefont {Fell}\ and\ \citenamefont {Axmacher}(2011)}]{FA11}%
  \BibitemOpen
  \bibfield  {author} {\bibinfo {author} {\bibfnamefont {J.}~\bibnamefont
  {Fell}}\ and\ \bibinfo {author} {\bibfnamefont {N.}~\bibnamefont
  {Axmacher}},\ }\bibfield  {title} {\enquote {\bibinfo {title} {The role of
  phase synchronization in memory processes},}\ }\href@noop {} {\bibfield
  {journal} {\bibinfo  {journal} {Nature Reviews Neuroscience}\ }\textbf
  {\bibinfo {volume} {12}},\ \bibinfo {pages} {105--118} (\bibinfo {year}
  {2011})}\BibitemShut {NoStop}%
\bibitem [{\citenamefont {Ott}\ and\ \citenamefont {Antonsen}(2008)}]{OA08}%
  \BibitemOpen
  \bibfield  {author} {\bibinfo {author} {\bibfnamefont {E.}~\bibnamefont
  {Ott}}\ and\ \bibinfo {author} {\bibfnamefont {T.~M.}\ \bibnamefont
  {Antonsen}},\ }\bibfield  {title} {\enquote {\bibinfo {title} {Low
  dimensional behavior of large systems of globally coupled oscillators},}\
  }\href {\doibase 10.1063/1.2930766} {\bibfield  {journal} {\bibinfo
  {journal} {Chaos}\ }\textbf {\bibinfo {volume} {18}},\ \bibinfo {eid}
  {037113} (\bibinfo {year} {2008})}\BibitemShut {NoStop}%
\bibitem [{\citenamefont {Hodgkin}(1948)}]{Hod48}%
  \BibitemOpen
  \bibfield  {author} {\bibinfo {author} {\bibfnamefont {AL}~\bibnamefont
  {Hodgkin}},\ }\bibfield  {title} {\enquote {\bibinfo {title} {The local
  electric changes associated with repetitive action in a non-medullated
  axon},}\ }\href@noop {} {\bibfield  {journal} {\bibinfo  {journal} {The
  Journal of physiology}\ }\textbf {\bibinfo {volume} {107}},\ \bibinfo {pages}
  {165--181} (\bibinfo {year} {1948})}\BibitemShut {NoStop}%
\bibitem [{\citenamefont {Rinzel}\ and\ \citenamefont
  {Ermentrout}(1998)}]{RE98}%
  \BibitemOpen
  \bibfield  {author} {\bibinfo {author} {\bibfnamefont {J.}~\bibnamefont
  {Rinzel}}\ and\ \bibinfo {author} {\bibfnamefont {B.}~\bibnamefont
  {Ermentrout}},\ }\bibfield  {title} {\enquote {\bibinfo {title} {Analysis of
  neural excitability and oscillations},}\ }in\ \href@noop {} {\emph {\bibinfo
  {booktitle} {Methods in Neuronal Modelling: From Ions to Networks}}},\
  \bibinfo {editor} {edited by\ \bibinfo {editor} {\bibfnamefont
  {C.}~\bibnamefont {Koch}}\ and\ \bibinfo {editor} {\bibfnamefont
  {I.}~\bibnamefont {Segev}}}\ (\bibinfo  {publisher} {MIT Press},\ \bibinfo
  {address} {Cambridge},\ \bibinfo {year} {1998})\ pp.\ \bibinfo {pages}
  {251--291}\BibitemShut {NoStop}%
\bibitem [{\citenamefont {Ermentrout}\ and\ \citenamefont
  {Kopell}(1986)}]{EK86}%
  \BibitemOpen
  \bibfield  {author} {\bibinfo {author} {\bibfnamefont {B.}~\bibnamefont
  {Ermentrout}}\ and\ \bibinfo {author} {\bibfnamefont {N.}~\bibnamefont
  {Kopell}},\ }\bibfield  {title} {\enquote {\bibinfo {title} {Parabolic
  bursting in an excitable system coupled with a slow oscillation},}\
  }\href@noop {} {\bibfield  {journal} {\bibinfo  {journal} {{SIAM} J. Appl.
  Math.}\ }\textbf {\bibinfo {volume} {46}},\ \bibinfo {pages} {233--253}
  (\bibinfo {year} {1986})}\BibitemShut {NoStop}%
\bibitem [{\citenamefont {Latham}\ \emph {et~al.}(2000)\citenamefont {Latham},
  \citenamefont {Richmond}, \citenamefont {Nelson},\ and\ \citenamefont
  {Nirenberg}}]{LRN+00}%
  \BibitemOpen
  \bibfield  {author} {\bibinfo {author} {\bibfnamefont {PE}~\bibnamefont
  {Latham}}, \bibinfo {author} {\bibfnamefont {BJ}~\bibnamefont {Richmond}},
  \bibinfo {author} {\bibfnamefont {PG}~\bibnamefont {Nelson}}, \ and\ \bibinfo
  {author} {\bibfnamefont {S}~\bibnamefont {Nirenberg}},\ }\bibfield  {title}
  {\enquote {\bibinfo {title} {Intrinsic dynamics in neuronal networks. i.
  theory},}\ }\href@noop {} {\bibfield  {journal} {\bibinfo  {journal} {Journal
  of Neurophysiology}\ }\textbf {\bibinfo {volume} {83}},\ \bibinfo {pages}
  {808--827} (\bibinfo {year} {2000})}\BibitemShut {NoStop}%
\bibitem [{\citenamefont {Buice}\ and\ \citenamefont {Chow}(2013)}]{bc13}%
  \BibitemOpen
  \bibfield  {author} {\bibinfo {author} {\bibfnamefont {M.~A.}\ \bibnamefont
  {Buice}}\ and\ \bibinfo {author} {\bibfnamefont {C.~C.}\ \bibnamefont
  {Chow}},\ }\bibfield  {title} {\enquote {\bibinfo {title} {Dynamic finite
  size effects in spiking neural networks},}\ }\href {\doibase
  10.1371/journal.pcbi.1002872} {\bibfield  {journal} {\bibinfo  {journal}
  {PLoS Comput. Biol.}\ }\textbf {\bibinfo {volume} {9}},\ \bibinfo {pages}
  {e1002872} (\bibinfo {year} {2013})}\BibitemShut {NoStop}%
\bibitem [{Note1()}]{Note1}%
  \BibitemOpen
  \bibinfo {note} {We make an analytic continuation of $w(\eta ,t)$ from real
  $\eta $ into complex-valued $\eta =\eta _r+i\eta _i$. This is possible into
  the lower half-plane $\eta _i<0$, since this guarantees the half-width
  $x(\eta ,t)$ remains positive zero: $\partial _t x(\eta ,t)=- \eta _i >0$ at
  $x=0$. Therefore we closed the integrals in \protect \textup {\hbox
  {\mathsurround \z@ \protect \normalfont (\ignorespaces \ref {r}\unskip
  \@@italiccorr )}} and \protect \textup {\hbox {\mathsurround \z@ \protect
  \normalfont (\ignorespaces \ref {v}\unskip \@@italiccorr )}} with an arc
  $|\eta | e^{i\vartheta }$ with $|\eta |\to \infty $ and $\vartheta \in (-\pi
  ,0)$. This contour encloses one pole of the Lorentzian distribution \protect
  \textup {\hbox {\mathsurround \z@ \protect \normalfont (\ignorespaces \ref
  {lorentzian}\unskip \@@italiccorr )}}, which written in partial fractions
  reads: $g(\eta )=(2\pi i)^{-1}[(\eta -\protect \mathaccentV {bar}016\eta
  -i\Delta )^{-1}- (\eta -\protect \mathaccentV {bar}016\eta +i\Delta
  )^{-1}]$.}\BibitemShut {Stop}%
\bibitem [{Note2()}]{Note2}%
  \BibitemOpen
  \bibinfo {note} {The number of effective parameters can be reduced by
  nondimensionalizing the system as: $\protect \mathaccentV {tilde}07E\eta =
  \protect \mathaccentV {bar}016\eta /\Delta $, $\protect \mathaccentV
  {tilde}07EJ= J/\protect \sqrt {\Delta }$, $(\protect \mathaccentV
  {tilde}07Er, \protect \mathaccentV {tilde}07Ev)=(r,v)/ \protect \sqrt {\Delta
  }$, $\protect \mathaccentV {tilde}07Et= \protect \sqrt \Delta t$, and
  $\protect \mathaccentV {tilde}07EI(\protect \mathaccentV
  {tilde}07Et)=I(t/\protect \sqrt \Delta )/\Delta $.}\BibitemShut {Stop}%
\bibitem [{\citenamefont {Kuramoto}(1984)}]{Kur84}%
  \BibitemOpen
  \bibfield  {author} {\bibinfo {author} {\bibfnamefont {Y.}~\bibnamefont
  {Kuramoto}},\ }\href@noop {} {\emph {\bibinfo {title} {Chemical Oscillations,
  Waves, and Turbulence}}}\ (\bibinfo  {publisher} {{S}pringer-{V}erlag},\
  \bibinfo {address} {Berlin},\ \bibinfo {year} {1984})\BibitemShut {NoStop}%
\bibitem [{\citenamefont {Pikovsky}\ \emph {et~al.}(2001)\citenamefont
  {Pikovsky}, \citenamefont {Rosenblum},\ and\ \citenamefont {Kurths}}]{PRK01}%
  \BibitemOpen
  \bibfield  {author} {\bibinfo {author} {\bibfnamefont {A.~S.}\ \bibnamefont
  {Pikovsky}}, \bibinfo {author} {\bibfnamefont {M.~G.}\ \bibnamefont
  {Rosenblum}}, \ and\ \bibinfo {author} {\bibfnamefont {J.}~\bibnamefont
  {Kurths}},\ }\href@noop {} {\emph {\bibinfo {title} {Synchronization, a
  Universal Concept in Nonlinear Sciences}}}\ (\bibinfo  {publisher} {Cambridge
  University Press},\ \bibinfo {address} {Cambridge},\ \bibinfo {year}
  {2001})\BibitemShut {NoStop}%
\bibitem [{\citenamefont {Ott}\ and\ \citenamefont {Antonsen}(2009)}]{OA09}%
  \BibitemOpen
  \bibfield  {author} {\bibinfo {author} {\bibfnamefont {E.}~\bibnamefont
  {Ott}}\ and\ \bibinfo {author} {\bibfnamefont {T.~M.}\ \bibnamefont
  {Antonsen}},\ }\bibfield  {title} {\enquote {\bibinfo {title} {Long time
  evolution of phase oscillator systems},}\ }\href {\doibase 10.1063/1.3136851}
  {\bibfield  {journal} {\bibinfo  {journal} {Chaos}\ }\textbf {\bibinfo
  {volume} {19}},\ \bibinfo {eid} {023117} (\bibinfo {year}
  {2009})}\BibitemShut {NoStop}%
\bibitem [{\citenamefont {Ott}\ \emph {et~al.}(2011)\citenamefont {Ott},
  \citenamefont {Hunt},\ and\ \citenamefont {Antonsen}}]{OHA11}%
  \BibitemOpen
  \bibfield  {author} {\bibinfo {author} {\bibfnamefont {E.}~\bibnamefont
  {Ott}}, \bibinfo {author} {\bibfnamefont {B.~R.}\ \bibnamefont {Hunt}}, \
  and\ \bibinfo {author} {\bibfnamefont {T.~M.}\ \bibnamefont {Antonsen}},\
  }\bibfield  {title} {\enquote {\bibinfo {title} {Comment on ``long time
  evolution of phase oscillators systems''},}\ }\href@noop {} {\bibfield
  {journal} {\bibinfo  {journal} {Chaos}\ }\textbf {\bibinfo {volume} {21}},\
  \bibinfo {eid} {025112} (\bibinfo {year} {2011})}\BibitemShut {NoStop}%
\bibitem [{\citenamefont {Paz\'o}\ and\ \citenamefont
  {Montbri\'o}(2014)}]{PM14}%
  \BibitemOpen
  \bibfield  {author} {\bibinfo {author} {\bibfnamefont {D.}~\bibnamefont
  {Paz\'o}}\ and\ \bibinfo {author} {\bibfnamefont {E.}~\bibnamefont
  {Montbri\'o}},\ }\bibfield  {title} {\enquote {\bibinfo {title}
  {Low-dimensional dynamics of populations of pulse-coupled oscillators},}\
  }\href {\doibase 10.1103/PhysRevX.4.011009} {\bibfield  {journal} {\bibinfo
  {journal} {Phys. Rev. X}\ }\textbf {\bibinfo {volume} {4}},\ \bibinfo {pages}
  {011009} (\bibinfo {year} {2014})}\BibitemShut {NoStop}%
\bibitem [{\citenamefont {Pikovsky}\ and\ \citenamefont
  {Rosenblum}(2011)}]{PR11}%
  \BibitemOpen
  \bibfield  {author} {\bibinfo {author} {\bibfnamefont {A.}~\bibnamefont
  {Pikovsky}}\ and\ \bibinfo {author} {\bibfnamefont {M.}~\bibnamefont
  {Rosenblum}},\ }\bibfield  {title} {\enquote {\bibinfo {title} {Dynamics of
  heterogeneous oscillator ensembles in terms of collective variables},}\
  }\href {\doibase DOI: 10.1016/j.physd.2011.01.002} {\bibfield  {journal}
  {\bibinfo  {journal} {Physica D}\ }\textbf {\bibinfo {volume} {240}},\
  \bibinfo {pages} {872 -- 881} (\bibinfo {year} {2011})}\BibitemShut {NoStop}%
\bibitem [{\citenamefont {Luke}\ \emph {et~al.}(2013)\citenamefont {Luke},
  \citenamefont {Barreto},\ and\ \citenamefont {So}}]{LBS13}%
  \BibitemOpen
  \bibfield  {author} {\bibinfo {author} {\bibfnamefont {T.~B.}\ \bibnamefont
  {Luke}}, \bibinfo {author} {\bibfnamefont {E.}~\bibnamefont {Barreto}}, \
  and\ \bibinfo {author} {\bibfnamefont {P.}~\bibnamefont {So}},\ }\bibfield
  {title} {\enquote {\bibinfo {title} {Complete classification of the
  macroscopic behavior of a heterogeneous network of theta neurons},}\
  }\href@noop {} {\bibfield  {journal} {\bibinfo  {journal} {Neural Comput.}\
  }\textbf {\bibinfo {volume} {25}},\ \bibinfo {pages} {3207--3234} (\bibinfo
  {year} {2013})}\BibitemShut {NoStop}%
\bibitem [{\citenamefont {So}\ \emph {et~al.}(2014)\citenamefont {So},
  \citenamefont {Luke},\ and\ \citenamefont {Barreto}}]{SLB14}%
  \BibitemOpen
  \bibfield  {author} {\bibinfo {author} {\bibfnamefont {P.}~\bibnamefont
  {So}}, \bibinfo {author} {\bibfnamefont {T.~B.}\ \bibnamefont {Luke}}, \ and\
  \bibinfo {author} {\bibfnamefont {E.}~\bibnamefont {Barreto}},\ }\bibfield
  {title} {\enquote {\bibinfo {title} {Networks of theta neurons with
  time-varying excitability: Macroscopic chaos, multistability, and final-state
  uncertainty},}\ }\href {\doibase
  http://dx.doi.org/10.1016/j.physd.2013.04.009} {\bibfield  {journal}
  {\bibinfo  {journal} {Physica D}\ }\textbf {\bibinfo {volume} {267}},\
  \bibinfo {pages} {16--26} (\bibinfo {year} {2014})}\BibitemShut {NoStop}%
\bibitem [{\citenamefont {Laing}(2014)}]{Lai14}%
  \BibitemOpen
  \bibfield  {author} {\bibinfo {author} {\bibfnamefont {C.~R.}\ \bibnamefont
  {Laing}},\ }\bibfield  {title} {\enquote {\bibinfo {title} {Derivation of a
  neural field model from a network of theta neurons},}\ }\href {\doibase
  10.1103/PhysRevE.90.010901} {\bibfield  {journal} {\bibinfo  {journal} {Phys.
  Rev. E}\ }\textbf {\bibinfo {volume} {90}},\ \bibinfo {pages} {010901}
  (\bibinfo {year} {2014})}\BibitemShut {NoStop}%
\bibitem [{\citenamefont {Montbri\'o}\ and\ \citenamefont
  {Paz\'o}(2011{\natexlab{a}})}]{MP11}%
  \BibitemOpen
  \bibfield  {author} {\bibinfo {author} {\bibfnamefont {E.}~\bibnamefont
  {Montbri\'o}}\ and\ \bibinfo {author} {\bibfnamefont {D.}~\bibnamefont
  {Paz\'o}},\ }\bibfield  {title} {\enquote {\bibinfo {title} {Shear diversity
  prevents collective synchronization},}\ }\href {\doibase
  10.1103/PhysRevLett.106.254101} {\bibfield  {journal} {\bibinfo  {journal}
  {Phys. Rev. Lett.}\ }\textbf {\bibinfo {volume} {106}},\ \bibinfo {pages}
  {254101} (\bibinfo {year} {2011}{\natexlab{a}})}\BibitemShut {NoStop}%
\bibitem [{\citenamefont {Paz{\'o}}\ and\ \citenamefont
  {Montbri{\'o}}(2011)}]{PM11}%
  \BibitemOpen
  \bibfield  {author} {\bibinfo {author} {\bibfnamefont {D.}~\bibnamefont
  {Paz{\'o}}}\ and\ \bibinfo {author} {\bibfnamefont {E.}~\bibnamefont
  {Montbri{\'o}}},\ }\bibfield  {title} {\enquote {\bibinfo {title} {The
  {K}uramoto model with distributed shear},}\ }\href {\doibase
  10.1209/0295-5075/95/60007} {\bibfield  {journal} {\bibinfo  {journal} {EPL
  (Europhys. Lett.)}\ }\textbf {\bibinfo {volume} {95}},\ \bibinfo {pages}
  {60007} (\bibinfo {year} {2011})}\BibitemShut {NoStop}%
\bibitem [{\citenamefont {Montbri\'o}\ and\ \citenamefont
  {Paz\'o}(2011{\natexlab{b}})}]{MP11p}%
  \BibitemOpen
  \bibfield  {author} {\bibinfo {author} {\bibfnamefont {E.}~\bibnamefont
  {Montbri\'o}}\ and\ \bibinfo {author} {\bibfnamefont {D.}~\bibnamefont
  {Paz\'o}},\ }\bibfield  {title} {\enquote {\bibinfo {title} {Collective
  synchronization in the presence of reactive coupling and shear diversity},}\
  }\href {\doibase 10.1103/PhysRevE.84.046206} {\bibfield  {journal} {\bibinfo
  {journal} {Phys. Rev. E}\ }\textbf {\bibinfo {volume} {84}},\ \bibinfo
  {pages} {046206} (\bibinfo {year} {2011}{\natexlab{b}})}\BibitemShut
  {NoStop}%
\bibitem [{\citenamefont {Iatsenko}\ \emph {et~al.}(2013)\citenamefont
  {Iatsenko}, \citenamefont {Petkoski}, \citenamefont {McClintock},\ and\
  \citenamefont {Stefanovska}}]{IPM+13}%
  \BibitemOpen
  \bibfield  {author} {\bibinfo {author} {\bibfnamefont {D.}~\bibnamefont
  {Iatsenko}}, \bibinfo {author} {\bibfnamefont {S.}~\bibnamefont {Petkoski}},
  \bibinfo {author} {\bibfnamefont {P.~V.~E.}\ \bibnamefont {McClintock}}, \
  and\ \bibinfo {author} {\bibfnamefont {A.}~\bibnamefont {Stefanovska}},\
  }\bibfield  {title} {\enquote {\bibinfo {title} {Stationary and traveling
  wave states of the {K}uramoto model with an arbitrary distribution of
  frequencies and coupling strengths},}\ }\href {\doibase
  10.1103/PhysRevLett.110.064101} {\bibfield  {journal} {\bibinfo  {journal}
  {Phys. Rev. Lett.}\ }\textbf {\bibinfo {volume} {110}},\ \bibinfo {pages}
  {064101} (\bibinfo {year} {2013})}\BibitemShut {NoStop}%
\bibitem [{\citenamefont {Iatsenko}\ \emph {et~al.}(2014)\citenamefont
  {Iatsenko}, \citenamefont {McClintock},\ and\ \citenamefont
  {Stefanovska}}]{IMS14}%
  \BibitemOpen
  \bibfield  {author} {\bibinfo {author} {\bibfnamefont {D.}~\bibnamefont
  {Iatsenko}}, \bibinfo {author} {\bibfnamefont {P.~V.~E.}\ \bibnamefont
  {McClintock}}, \ and\ \bibinfo {author} {\bibfnamefont {A.}~\bibnamefont
  {Stefanovska}},\ }\bibfield  {title} {\enquote {\bibinfo {title} {Oscillator
  glass in the generalized {K}uramoto model: synchronous disorder and two-step
  relaxation},}\ }\href {\doibase 10.1038/ncomms5118} {\bibfield  {journal}
  {\bibinfo  {journal} {Nat. Commun.}\ }\textbf {\bibinfo {volume} {5}},\
  \bibinfo {pages} {4188} (\bibinfo {year} {2014})}\BibitemShut {NoStop}%
\bibitem [{Note3()}]{Note3}%
  \BibitemOpen
  \bibinfo {note} {However, note that the phase diagrams in Luke et al. \cite
  {LBS13} do not fully agree with ours, and show additional instabilities.
  These bifurcations are associated with the finite width of the pulses, as it
  accurs in the Winfree model \cite {PM14}.}\BibitemShut {Stop}%
\bibitem [{\citenamefont {Roberts}\ and\ \citenamefont {Quispel}(1992)}]{rq92}%
  \BibitemOpen
  \bibfield  {author} {\bibinfo {author} {\bibfnamefont {J.A.G.}\ \bibnamefont
  {Roberts}}\ and\ \bibinfo {author} {\bibfnamefont {G.R.W.}\ \bibnamefont
  {Quispel}},\ }\bibfield  {title} {\enquote {\bibinfo {title} {Chaos and
  time-reversal symmetry. {O}rder and chaos in reversible dynamical systems},}\
  }\href {\doibase http://dx.doi.org/10.1016/0370-1573(92)90163-T} {\bibfield
  {journal} {\bibinfo  {journal} {Phys. Rep.}\ }\textbf {\bibinfo {volume}
  {216}},\ \bibinfo {pages} {63 -- 177} (\bibinfo {year} {1992})}\BibitemShut
  {NoStop}%
\bibitem [{\citenamefont {Strogatz}\ and\ \citenamefont
  {Mirollo}(1991)}]{SM91}%
  \BibitemOpen
  \bibfield  {author} {\bibinfo {author} {\bibfnamefont {S.~H.}\ \bibnamefont
  {Strogatz}}\ and\ \bibinfo {author} {\bibfnamefont {R.~E.}\ \bibnamefont
  {Mirollo}},\ }\bibfield  {title} {\enquote {\bibinfo {title} {Stability of
  incoherence in a population of coupled oscillators},}\ }\href@noop {}
  {\bibfield  {journal} {\bibinfo  {journal} {J. Stat. Phys.}\ }\textbf
  {\bibinfo {volume} {63}},\ \bibinfo {pages} {613--635} (\bibinfo {year}
  {1991})}\BibitemShut {NoStop}%
\bibitem [{\citenamefont {Strogatz}\ \emph {et~al.}(1992)\citenamefont
  {Strogatz}, \citenamefont {Mirollo},\ and\ \citenamefont {Matthews}}]{SMM92}%
  \BibitemOpen
  \bibfield  {author} {\bibinfo {author} {\bibfnamefont {S.~H.}\ \bibnamefont
  {Strogatz}}, \bibinfo {author} {\bibfnamefont {R.~E.}\ \bibnamefont
  {Mirollo}}, \ and\ \bibinfo {author} {\bibfnamefont {P.~C.}\ \bibnamefont
  {Matthews}},\ }\bibfield  {title} {\enquote {\bibinfo {title} {Coupled
  nonlinear oscillators below the synchronization threshold: Relaxation by
  generalized {L}andau damping},}\ }\href {\doibase
  10.1103/PhysRevLett.68.2730} {\bibfield  {journal} {\bibinfo  {journal}
  {Phys. Rev. Lett.}\ }\textbf {\bibinfo {volume} {68}},\ \bibinfo {pages}
  {2730--2733} (\bibinfo {year} {1992})}\BibitemShut {NoStop}%
\bibitem [{\citenamefont {Chiba}\ and\ \citenamefont
  {Nishikawa}(2011)}]{chiba11}%
  \BibitemOpen
  \bibfield  {author} {\bibinfo {author} {\bibfnamefont {H.}~\bibnamefont
  {Chiba}}\ and\ \bibinfo {author} {\bibfnamefont {I.}~\bibnamefont
  {Nishikawa}},\ }\bibfield  {title} {\enquote {\bibinfo {title} {Center
  manifold reduction for large populations of globally coupled phase
  oscillators},}\ }\href {\doibase http://dx.doi.org/10.1063/1.3647317}
  {\bibfield  {journal} {\bibinfo  {journal} {Chaos}\ }\textbf {\bibinfo
  {volume} {21}},\ \bibinfo {eid} {043103} (\bibinfo {year}
  {2011})}\BibitemShut {NoStop}%
\bibitem [{\citenamefont {Daido}(1996)}]{Dai96}%
  \BibitemOpen
  \bibfield  {author} {\bibinfo {author} {\bibfnamefont {H.}~\bibnamefont
  {Daido}},\ }\bibfield  {title} {\enquote {\bibinfo {title} {Onset of
  cooperative entrainment in limit-cycle oscillators with uniform all-to-all
  interactions: bifurcation of the order function},}\ }\href@noop {} {\bibfield
   {journal} {\bibinfo  {journal} {Physica D}\ }\textbf {\bibinfo {volume}
  {91}},\ \bibinfo {pages} {24--66} (\bibinfo {year} {1996})}\BibitemShut
  {NoStop}%
\bibitem [{Note4()}]{Note4}%
  \BibitemOpen
  \bibinfo {note} {Equation~\protect \textup {\hbox {\mathsurround \z@ \protect
  \normalfont (\ignorespaces \ref {w}\unskip \@@italiccorr )}} becomes then a
  reversible system \cite {rq92} of ordinary differential equations with the
  invariance under time reversal and $w\to w^*$, e.g.~Eq.~\protect \textup
  {\hbox {\mathsurround \z@ \protect \normalfont (\ignorespaces \ref
  {fre}\unskip \@@italiccorr )}} for $\Delta =0$ is reversible, with invariance
  under $t\to -t$, $v\to -v$. It is advisable to approximate Eq.~\protect
  \textup {\hbox {\mathsurround \z@ \protect \normalfont (\ignorespaces \ref
  {w}\unskip \@@italiccorr )}} by a set of ODEs (one for each $\eta $ value)
  with additional small terms breaking time-reverisibility. This is achieved
  including a small term in each ODE, like $\Delta $ in Eq.~\protect \textup
  {\hbox {\mathsurround \z@ \protect \normalfont (\ignorespaces \ref
  {frea}\unskip \@@italiccorr )}}, so that $g(\eta )$ is actually approximated
  by a sum of very sharp Lorentzian functions.}\BibitemShut {Stop}%
\bibitem [{\citenamefont {Watanabe}\ and\ \citenamefont
  {Strogatz}(1994)}]{WS94}%
  \BibitemOpen
  \bibfield  {author} {\bibinfo {author} {\bibfnamefont {S.}~\bibnamefont
  {Watanabe}}\ and\ \bibinfo {author} {\bibfnamefont {S.~H.}\ \bibnamefont
  {Strogatz}},\ }\bibfield  {title} {\enquote {\bibinfo {title} {Constant of
  motion for superconducting {J}osephson arrays},}\ }\href@noop {} {\bibfield
  {journal} {\bibinfo  {journal} {Physica D}\ }\textbf {\bibinfo {volume}
  {74}},\ \bibinfo {pages} {197--253} (\bibinfo {year} {1994})}\BibitemShut
  {NoStop}%
\bibitem [{\citenamefont {Pikovsky}\ and\ \citenamefont
  {Rosenblum}(2008)}]{PR08}%
  \BibitemOpen
  \bibfield  {author} {\bibinfo {author} {\bibfnamefont {A.}~\bibnamefont
  {Pikovsky}}\ and\ \bibinfo {author} {\bibfnamefont {M.}~\bibnamefont
  {Rosenblum}},\ }\bibfield  {title} {\enquote {\bibinfo {title} {Partially
  integrable dynamics of hierarchical populations of coupled oscillators},}\
  }\href {\doibase 10.1103/PhysRevLett.101.264103} {\bibfield  {journal}
  {\bibinfo  {journal} {Phys. Rev. Lett.}\ }\textbf {\bibinfo {volume} {101}},\
  \bibinfo {pages} {264103} (\bibinfo {year} {2008})}\BibitemShut {NoStop}%
\bibitem [{Note5()}]{Note5}%
  \BibitemOpen
  \bibinfo {note} {A subtle question is that, as it occurs in the Kuramoto
  model, there exists a continuous spectrum of eigenvalues at $\lambda
  =-2w_0(\eta )i$ and the complex conjugate. Indeed, for $\eta $ values such
  that neurons are actively firing ($y_0(\eta )=0$), the eigenvalues lie
  exactly on the imaginary axis. Apparently, this seems to forbid the
  exponential stability obtained from Eq.~\protect \textup {\hbox
  {\mathsurround \z@ \protect \normalfont (\ignorespaces \ref {fre}\unskip
  \@@italiccorr )}} for the Lorentzian distribution. The same mathematical
  conundrum was encountered and solved in the Kuramoto model \cite
  {SM91,SMM92,chiba11}, what suggests similar mechanisms are operating
  here.}\BibitemShut {Stop}%
\bibitem [{Note6()}]{Note6}%
  \BibitemOpen
  \bibinfo {note} {Note that considering stronger forms of heterogeneity for
  the synaptic weights may generally not lead to Eq.~\protect \textup {\hbox
  {\mathsurround \z@ \protect \normalfont (\ignorespaces \ref
  {general_QIF}\unskip \@@italiccorr )}}.}\BibitemShut {Stop}%
\end{thebibliography}

%

\newpage
\section*{Appendix A: Numerical Simulations }
To numerically simulate the population of QIF neurons \eqref{qif} we used 
the Euler method, with time step $dt=10^{-4}$. 
The population had $N=10^4$ neurons, and the 
Lorentzian distribution  
\eqref{lorentzian} was deterministically 
generated using: $\eta_j= \bar\eta +\Delta \arctan[\pi/2 (2j-N-1)/(N+1)]$, 
where $j=1,\dots,N$, and $\Delta=1$. 

The time it takes for the membrane potential of a QIF neuron 
(with $I_j>0$) to reach infinity from a given positive value of the 
membrane potential ̇is $\arctan(\sqrt{I_j}/V_p)/\sqrt{I_j}$. 
For $\sqrt{I_j}\ll V_p$,  this expression can be 
approximated as: $$\frac{\arctan\left(\sqrt{I_j}/V_p\right)}{\sqrt{I_j}} 
\approx \frac{1}{V_p} .$$

In simulations we considered $V_p=-V_r=100$, and used the previous 
approximation. Thus, the time for the neurons to reach infinity from $V_p$
is $1/V_p \approx 10^{-2}$. Additionally, the time taken from minus infinity 
to $V_r$ is $−1/V_r \approx 10^{-2}$. Numerically, once a neuron's 
membrane potential $V_j$  satisfies $V_j \geq V_p$ the neuron is 
reset to $-V_j$ and held there for a refractory time $2/V_j$. 
The neurons produce a spike when 
$V_j$ reaches infinity, i.e.~a time interval $1/V_j$ after crossing $V_p$.
The exact time of the spike can not be evaluated exactly 
in numerical simulations, 
since $dt$ is finite. However, simulations agree very well with the 
theory provided that $1/V_p \gg dt$.   

To evaluate the mean membrane potential $v$, 
the population average was computed discarding those neurons
in the refractory period. The choice $V_p=-V_r$ is not 
needed, but in this way the observed mean membrane potential $v$ agrees with the
theory, consistent with the definition in \eqref{y}.
The mean synaptic activation \eqref{s} was evaluated 
using the Heaviside step function $a_\tau(t)=\Theta(\tau-t)/\tau$ with 
$\tau=10^{-3}$. 
The instantaneous firing rates were obtained binning 
time and counting spikes within a sliding time window of size
$\delta t=2\times10^{-2}$ 
(Figs.~\ref{Fig2},~\ref{FigA2},~\ref{FigA3},~\ref{FigA4}) 
and $\delta t=4\times10^{-2}$ (Fig.~\ref{Fig3}).

 \makeatletter 
\setcounter{equation}{0} 
\renewcommand{\theequation}{B\arabic{equation}}
 \makeatother

\section*{Appendix B: Proof of equation \eqref{W}}

The inverse of Eq.~\eqref{W} reads:
\begin{equation}
W=\frac{1-Z^*}{1+Z^*}
\label{wz}
\end{equation}
Recalling Eqs.~\eqref{r} and \eqref{v}, we write the macroscopic field $W$ as
\begin{equation}
W(t)=\int_{-\infty}^\infty w(\eta,t) g(\eta) d\eta .\nonumber
\end{equation}
where $w\equiv x+iy$. Inserting the conformal mapping  
$w=(1-\alpha)/(1+\alpha)$ ---the inverse of Eq.~\eqref{alpha}-- we get
\begin{equation}
W(t)=\int_{-\infty}^\infty \left[ \frac{1-\alpha(\eta,t)}{1+\alpha(\eta,t)} \right]g(\eta) d\eta .\nonumber
\end{equation}
Using the geometric series formula,
and grouping the powers of $\alpha$, we may evaluate the integrals obtaining
\begin{equation}
W(t)=1-2 Z^*(t) +2 Z_2^*(t) - 2 Z_3^*(t) + \cdots , \label{wzm} 
\end{equation}
where the $Z_m$ are the generalized order parameters \cite{Dai96}:
\begin{eqnarray}
Z_m(t) &=&\int_{-\infty}^{\infty} g(\eta)\int_{0}^{2\pi} \tilde\rho(\theta|\eta,t) e^{im\theta}
d\theta \, d\eta  \nonumber\\&=&\int_{-\infty}^{\infty} g(\eta) [\alpha^*(\eta,t)]^m \,d\eta . \nonumber
\end{eqnarray}
Since, for Lorentzian $g(\eta)$, $Z_m(t)=[\alpha^*(\eta=\bar\eta-i\Delta,t)]^m=Z(t)^m$,
we can revert the power series in \eqref{wzm} and obtain the result in \eqref{wz}.

 \makeatletter 
\setcounter{equation}{0} 
\setcounter{figure}{0} 
\renewcommand{\theequation}{C\arabic{equation}}
\renewcommand{\thefigure}{C\@arabic\c@figure}
 \makeatother

\section*{Appendix C: Results for arbitrary distributions of currents}

In this Appendix we compare the results in the main text, obtained 
using a Lorentzian distribution of currents $g(\eta)$, with other
distributions. The results are qualitatively similar as evidenced 
by Figs.~\ref{FigA1},~\ref{FigA2} and \ref{FigA3} 
(cf.~Figs.~1, 2, and 3 in the main text).

Note that 
if a particular distribution $g(\eta)$ has
$2n$ poles (all of them off the real axis) one can readily obtain the FREs 
consisting of $n$ complex-valued ordinary differential equations. Even if $g(\eta)$ does not fulfill 
this condition,
as in the case of Gaussian or uniform distributions, 
the Lorentzian Ansatz still holds (see below).

Moreover, it is possible to efficiently simulate the dynamics of a population
by integrating Eq.~\eqref{w} for a sample of $\eta$ values that
approximate a particular $g(\eta)~ $\footnote{Equation~\eqref{w}
becomes then a reversible system 
\cite{rq92} of ordinary differential equations with the invariance under time reversal and $w\to w^*$, 
e.g.~Eq.~\eqref{fre} for $\Delta=0$ is reversible, with invariance under $t\to-t$, $v\to-v$.
It is advisable to approximate Eq.~\eqref{w} by  
a set of ODEs (one for each $\eta$ value) with additional small
terms breaking time-reverisibility. This is achieved including a small
term in each ODE, like $\Delta$ in Eq.~\eqref{frea}, so that $g(\eta)$ is actually
approximated by a sum of very sharp Lorentzian functions.}.
However, if $g(\eta)$ is a discrete distribution (or has a discrete part), 
---as in the case of the Dirac delta function 
$g(\eta)=\delta(\eta-\bar\eta)$---, the so-called Watanabe-Strogatz theory 
applies \cite{WS94,PR11,PR08}, and the `Lorentzian manifold' is no longer 
attracting since the dynamics is extremely degenerate, akin to a Hamiltonian system.

\subsection*{Steady states}

Considering $I(t)=0$ in Eqs.~\eqref{qif} and \eqref{qif2}, it is possible to find the 
solutions with steady firing rate $r(t)=r_0$ 
for arbitrary distributions of currents. 

As we pointed out in the main text, the stationary solution of the 
continuity equation \eqref{cont} for those neurons that are intrinsically active 
($\eta>-Jr_0$) must be is inversely proportional to their speed, that is 
\begin{equation}
\rho_0(V|\eta)= \frac{c(\eta)}{V^2+\eta+J r_0}, \nonumber
\end{equation}
where the normalizing constant is $c(\eta)=\sqrt{\eta+Jr_0} /\pi$.
The firing rate for each value of $\eta$ is then
$$r_0(\eta)=\rho_0(V\to\infty|\eta) \dot V(V\to\infty|\eta,t)=c(\eta),$$
and integration over all the firing neurons, gives the self-consistent condition
for $r_0$:
\begin{equation}\label{sc}
r_0 =\int_{-J r_0}^{\infty} \, c(\eta) g(\eta) \, d\eta 
\end{equation}
which is valid for any distribution $g(\eta)$ (an analogous expression 
was obtained in Eq.~(45) of \cite{bc13}).
For Lorentzian $g(\eta)$, the integral in \eqref{sc} can be evaluated, 
and solving the resulting equation for $r_0$ one obtains the steady 
states in agreement
with the result of setting $\dot r=\dot v=0$ in the FREs \eqref{fre}.

\subsection*{Linear stability analysis of steady states}
Assuming that the LA captures the actual dynamics of the population 
of QIF neurons, one can also investigate the stability of steady states 
for arbitrary distributions $g(\eta)$. Indeed, 
the evolution of infinitesimal perturbations from a steady state
\begin{equation}
 w_0(\eta)\equiv x_0(\eta)+i y_0(\eta)=
  \begin{cases}
\sqrt{\eta+J r_0} & \text{if } \eta>-J r_0 \\
-i\sqrt{-\eta-J r_0}    & \text{if } \eta\le-J r_0
  \end{cases}
\nonumber
\end{equation}
is determined by the linearization of Eq.~\eqref{w}:
\begin{equation}\label{dw}
\partial_t  \delta w(\eta,t) = i (J \delta r - 2 w_0 \delta w) 
\end{equation}
where $ \delta r(t)= ( 2\pi)^{-1} \int_{-\infty}^\infty [\delta w(\eta,t) + \delta w^*(\eta,t)] g(\eta) d\eta$.

To find the discrete spectrum of eigenvalues
~\footnote{A subtle question is that, as it occurs in the Kuramoto model,
there exists a continuous spectrum of eigenvalues at $\lambda=-2w_0(\eta)i$ and 
the complex conjugate. Indeed, for $\eta$ values such that neurons are 
actively firing ($y_0(\eta)=0$),
the eigenvalues lie exactly on the imaginary axis. Apparently, this seems to 
forbid the exponential stability obtained from Eq.~\eqref{fre} for the Lorentzian 
distribution. The same mathematical conundrum was
encountered and solved in the Kuramoto model \cite{SM91,SMM92,chiba11},
what suggests similar mechanisms are operating here.}, 
we use the ansatz $\delta w (\eta,t)= b(\eta) e^{\lambda t}$ in \eqref{dw} and obtain
\begin{equation}
[\lambda +2 w_0(\eta) i] b(\eta) e^{\lambda t} =
\frac{iJ}{2\pi} \int _{-\infty}^\infty [b(\eta) e^{\lambda t} +b^*(\eta) e^{\lambda^* t}] g(\eta) d\eta.  \nonumber
\end{equation}
Solving the equation for $be^{\lambda t}$, we find
\begin{equation}
b(\eta) e^{\lambda t}=\frac{1}{2 w_0(\eta)-i\lambda}
\frac{J}{2\pi} \int_{-\infty}^\infty [b(\eta)e^{\lambda t}+b^*(\eta)e^{\lambda^*t} ] g(\eta) d\eta.  \nonumber
\end{equation}
After summing the complex conjugate at each side of the equation,
we multiply by $g(\eta)$ and integrate over $\eta$. This allows one to cancel out
the integrals $\int(be^{\lambda t}+b^* e^{\lambda^* t}) g d\eta$. Then, imposing 
self-consistency, we find
\begin{equation}\label{sc2}
1= \frac{J}{2\pi} \int_{-\infty}^\infty \left[\frac{1}{2 w_0(\eta)-i\lambda} + \mbox{c.c.}\right] g(\eta) d\eta. 
\end{equation}
Local bifurcations are associated with marginal stability of the fixed points:
$\mathrm{Re}(\lambda)\to0$.
In correspondence with the results obtained for a Lorentzian distribution
in the main text, we expect bifurcations of saddle-node type associated with 
$\lambda=0$. Therefore, from Eq.~\eqref{sc2} we find that the loci of the 
saddle-node bifurcations are located at
\begin{equation}\label{jsnn}
 J_{SN}=\frac{2\pi}{\int_{-\infty}^\infty \frac{x_0(\eta)}{|w_0(\eta)|^2}  g(\eta) d\eta }. 
\end{equation}
Since $x_0=\sqrt{\eta+J r_0}$ for $\eta> -J r_0$, and zero otherwise, we can restrict the integration over $\eta$ to the range $(-J r_0,\infty)$. 
Now multiplying Eq.~\eqref{jsnn} by Eq.~\eqref{sc} 
we get, after defining $\xi=J_{SN}r_0$, 
two equations that permit to find the locus of the saddle-node bifurcations
systematically for arbitrary distributions of currents:
\begin{equation} \label{jsn}
 J_{SN}=\frac{2\pi}{\int_0^\infty \eta^{-1/2} g(\eta-\xi) d\eta} ,
\end{equation}
with $\xi$ obtained solving
\begin{equation}\label{xi}
 \xi=\frac{2 \int_0^\infty \eta^{1/2} g(\eta-\xi) d\eta}{\int_0^\infty \eta^{-1/2} g(\eta-\xi) d\eta}.
\end{equation}

\begin{figure}
\includegraphics[width=80mm,clip=true]{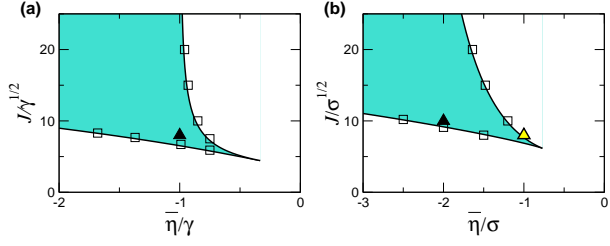}
\caption{(color online). Phase diagrams for (a) uniform and (b) Gaussian distributions of 
currents $g(\eta)$. The cyan-shaded region represents the bistable region,   
with the solid lines corresponding to saddle-node bifurcations analytically obtained 
from Eqs.~\eqref{jsn} and \eqref{xi}.
Square symbols are estimations of the bifurcations loci 
obtained from by direct numerical simulations
of $N=10^4$ QIF neurons. Triangle symbols indicate 
parameter values used in numerical simulations of 
Figs.~\ref{FigA2} and \ref{FigA3}.}
\label{FigA1}
\end{figure}

\begin{figure*}
\includegraphics[width=160mm,clip=true]{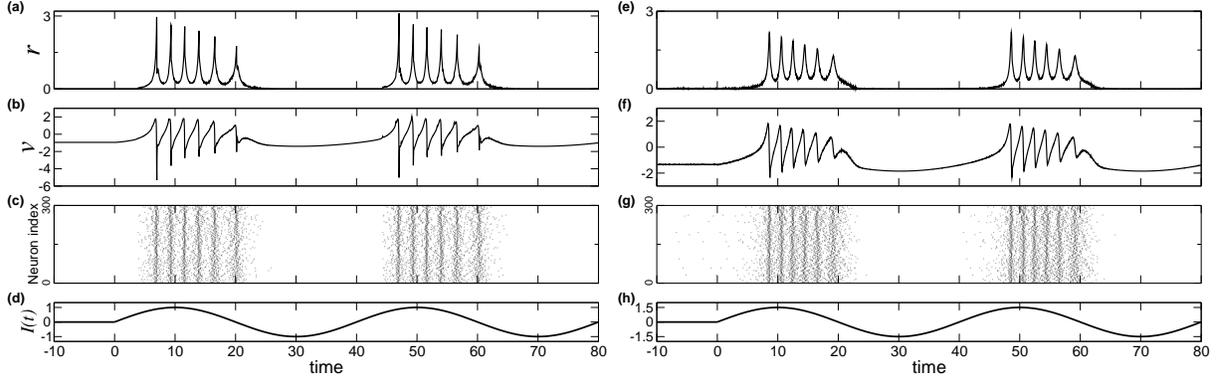}
\caption{Numerical simulations of the excitatory network of 
QIF neurons, Eqs.~\eqref{qif}, and \eqref{qif2}. 
with (a-d) uniform and (e-h) Gaussian distributions of currents. 
As in Fig.~\ref{Fig2}, at time $t=0$, 
an external sinusoidal current $I(t)=I_0 \sin(\omega t)$  ---shown in panels (d,h)---
is applied to all neurons. Panels (a,e) show the time series of the 
firing rate, and (b,f) the mean membrane potential of all cells. In panels 
(c,g) we depicted the raster plots of 300 randomly chosen neurons. 
Parameters correspond to the black triangle symbols in  
Fig.~\ref{FigA1}: (a-d) $\bar \eta=-1$, $J=8$, $\gamma=1$, $I_0=1$; (e-h) 
$\bar \eta=-2$, $J=10$, $\sigma=1$, $I_0=1.5$.}
\label{FigA2}
\end{figure*}

\subsection*{Bistability region for uniform and Gaussian distributions}

Equations \eqref{jsn} and \eqref{xi} are particularly easy to solve 
for uniform distributions of currents:
\begin{equation}
g(\eta)=\begin{cases} 

\frac{1}{2 \gamma}  & \mbox{for} \,  |\eta-\bar\eta|<\gamma 
\\
0  & \mbox{otherwise}
\end{cases} \nonumber
\end{equation}
After defining the rescaled parameters $\tilde\eta=\bar\eta/\gamma$ and
$\tilde J=  J/\sqrt{\gamma}$, we find two branches of saddle-node bifurcations
emanating from the cusp point at $\tilde\eta=-1/3$:
\begin{eqnarray}
\tilde J_{SN}^{(1)} &=& \frac{2\pi}{\sqrt{3 \tilde \eta+3}} \nonumber \\ 
\tilde J_{SN}^{(2)} &=& \frac{2\pi}{
\sqrt{\tilde\eta+1+2  \sqrt{\frac{1}{3}+\tilde\eta^2} } -   
\sqrt{\tilde\eta-1+2  \sqrt{\frac{1}{3}+\tilde\eta^2} } } \nonumber
\end{eqnarray}
These functions are plotted in Fig.~S1(a). Numerical simulations 
using the network of QIF neurons confirm the correctness of these
boundaries (see the square symbols in Fig.~\ref{FigA1}).

For Gaussian distributions, 
$$
g(\eta)= \frac{1}{\sqrt{2\pi} \sigma} e^{-(\eta-\bar\eta)^2/(2\sigma^2)},
$$
the solutions of \eqref{jsn} and \eqref{xi} can be numerically found. 
The results are shown in Fig.~\ref{FigA1}(b). Again, there
is a perfect agreement between Eqs.~\eqref{jsn} and \eqref{xi}
---derived assuming the Lorentzian Ansatz--- and the numerical estimations
obtained simulating the network of QIF neurons.

\subsection*{Excitatory networks with external periodic currents}
Firing rate equations \eqref{fre} predict the existence of a stable focus in 
the shaded region of Fig.~1. Trajectories attracted to this fixed point
display oscillations in the firing rate and mean membrane potential due to
the transient synchronous firing of the QIF neurons.  

When an external periodic current is injected to all neurons in the network, 
the spiral dynamics around the fixed point is responsible
for the bursting behavior observed in Fig.~2, as well as for
the emergence of the macroscopic chaos shown in Fig.~3. 
Here we investigate whether similar phenomena occur when an external periodic current 
of the same frequency is injected in an excitatory network with either uniform or 
Gaussian-distributed currents.

We introduced a sinusoidal forcing $I(t)=I_0 \sin(\omega t)$ 
observing a behavior qualitatively identical to the one reported in the main 
text with Lorentzian $g(\eta)$. Under a low-frequency forcing the system displays 
periodic bursting, see Fig.~\ref{FigA2}, provided the parameters are set inside the bistable region 
(see the black triangles in Fig.~\ref{FigA1}). Furthermore, note that the 
range of firing rates and mean membrane potentials in Figure~\ref{FigA2} are similar to 
those of Fig.~\ref{Fig2} ---for Lorentzian distributions of currents. 

\begin{figure}
\includegraphics[width=65mm,clip=true]{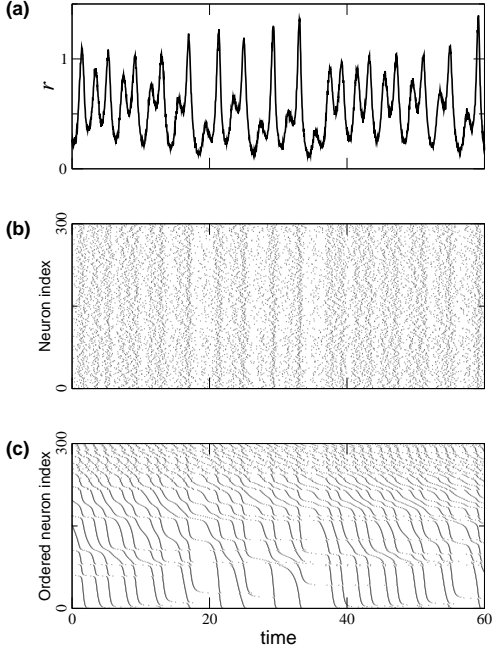}
\caption{Chaotic state in a network of $N=10^4$ QIF neurons
with Gaussian-distributed currents.
Network parameters correspond to the yellow-triangle symbol 
in the phase diagram of Fig.~S1(b). Like in Fig.~\ref{Fig3}, neurons receive 
a common periodic current $I(t)=I_0 \sin(\omega t)$ of frequency $\omega=\pi$.  
Parameters: $\bar \eta=-1$, $J=8$, $\sigma=1$, $I_0=1.5$.}
\label{FigA3}
\end{figure}

As for the simulations shown in Figure \eqref{Fig3}, 
we next increase the frequency of 
the injected current up to $\omega=\pi$ to investigate the 
emergence of macroscopic chaos in a network with Gaussian-distributed currents. 
Fig.~\ref{FigA3} shows the emergence of an apparently chaotic state. 
The observed dynamics
is similar to that of Fig.~\ref{Fig3}, which suggests it is chaotic. 
This type of chaos persists in the thermodynamic 
limit $N\to\infty$.  On top of this, highly-dimensional but weakly chaotic 
dynamics is probably also present due to `residual' finite-size effects.

 \makeatletter 
\setcounter{equation}{0} 
\setcounter{figure}{0} 
\renewcommand{\theequation}{D\arabic{equation}}
\renewcommand{\thefigure}{D\@arabic\c@figure}
 \makeatother

\section*{Appendix D: Identical QIF neurons driven by independent noise terms}
We compare now the results obtained above and in the main text 
for quenched heterogeneity $g(\eta)$ with the results for identical neurons
$g(\eta)=\delta(\eta-\bar\eta)$ driven by noise.
Specifically, the inputs currents are now taken as
\begin{equation}\label{noise}
 I_j=\bar\eta + J r(t) + \xi_j(t),
\end{equation}
where $\xi_j$ are independent white noise terms with expected values
$\left< \xi_j(t)\right>=0$, and $\left< \xi_j(t) \xi_k(t') \right>=
2 D \, \delta_{jk} \, \delta(t-t')$.
%
\begin{figure*}
\centerline{\includegraphics[width=160mm,clip=true]{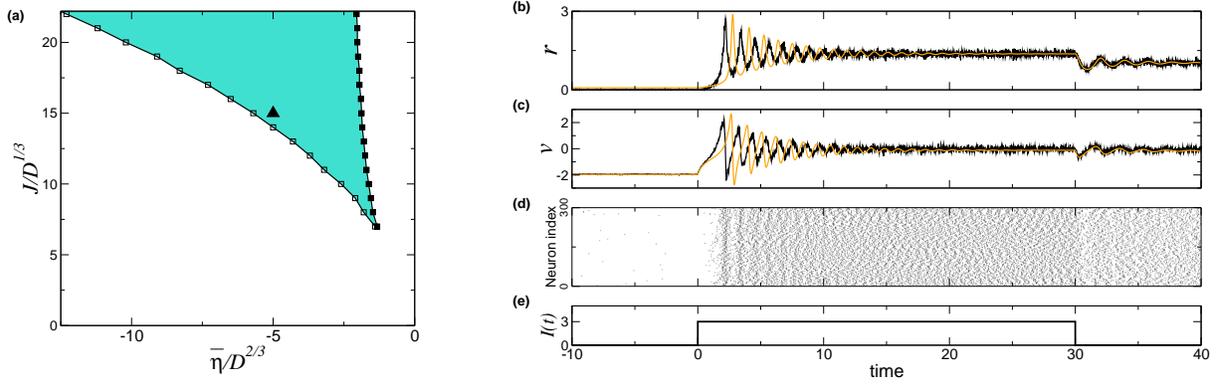}}
\caption{(color online). Numerically obtained phase diagram (a) and time series (b-d) 
for a network of identical QIF neurons driven by independent white noise terms. 
The square symbols in panel (a) represent the 
boundaries that enclose the region of bistability between a high and a 
low activity state. The boundaries have been obtained using a network of 
$10^4$ neurons, and by numerical 
continuation of a low activity solution (filled squares) and a high activity solution 
(empty squares). Specifically, using a noise intensity $D=1$ and a particular 
value of $J$, the system was initialized 
either at $\bar \eta=-5$ or at $\bar \eta=0$ and, after $t=10$ time units, 
parameter $\bar \eta$ was decreased/increased an amount $0.025/0.1$, respectively. This 
continuation was made until the relative change of two successive values of the 
averaged firing rate (time-averaged over the last time unit for each parameter value)
was larger than $50\%$. 
Other parameters for the numerical simulations are the same as in all other figures, 
and are described in the Material and Methods section. 
The triangle symbol indicates the parameter value, 
$(\bar \eta/D^{2/3},J/D^{1/3})=(-5,15)$,
corresponding to the numerical simulations shown in black 
in the right panels ---for the simulations we used $D=1$.
To facilitate the comparison with the FREs \eqref{fre}, 
the orange curves show the time series of the FREs, using the same
parameters $(\bar \eta,J)=(-5,15)$, but with $\Delta=1$} 
\label{FigA4}
\end{figure*}

In the thermodynamic limit, the density $\rho(V,t)$ obeys the Fokker-Planck equation:
\begin{equation}
\partial_t \rho + \partial_V\left[  (V^2+\bar\eta+J r) \rho \right]=D \,\partial^2_{V} \rho .
\label{fp}
\end{equation}
The Lorentzian ansatz is not a solution of this equation, but we demonstrate here
that the phenomena observed with quenched heterogeneity arise also
with independent noise sources. This qualitative similarity at the 
macroscopic level between quenched Lorentzian heterogeneity and Gaussian noise 
has been noted in previous work \cite{SM91}.  

A subtle point in Eq.~\eqref{fp} is that its nondimensionalization
entails a different (compared to $\Delta$) rescaling with $D$: $\tilde V=V/D^{1/3}$,
$\tilde \eta=\eta/D^{2/3}$, $\tilde J= J/D^{1/3}$, $\tilde t=t D^{1/3}$
(implying $\tilde r= r/D^{1/3}$).

Numerical simulations reveal the existence of a region
of bistability, see Fig.~\ref{FigA4}(a), analogous to the one observed for quenched heterogeneity,
cf.~Figs.~\ref{Fig1}(a) and \ref{FigA1}. 
Obviously, true bistability only holds in the thermodynamic
limit, while what we observe are rather exceedingly long
residence times close to each fixed point (see figure caption for details).

Additionally, in order to investigate and compare the
dynamical behavior of the identical noise-driven neurons 
with that of the FREs \eqref{fre}, 
an external current of intensity $I_0=3$ is injected to all neurons
at time $t=0$, like in Fig.~\ref{Fig2}. 
In Fig.~\ref{FigA4}(b,c) the time series of the firing rate and the mean membrane potential
clearly display damped oscillations after the injection of the current,
confirming the existence of a stable focus, exactly as observed in 
the FREs \eqref{fre}. The existence of a stable focus reflects the presence of 
transient spike synchrony in the network, as seen by the 
raster plot in Fig.~\ref{FigA4}(d). Remarkably, the raster plot and the time series closely 
resemble those of Fig.~\ref{Fig2}. The resemblance is not just qualitative, but rather there is 
a near quantitative fit between the network of QIF neurons driven by Gaussian noise 
and the FREs \eqref{fre}, which were derived assuming quenched Lorentzian 
heterogeneity (orange curves).

 \makeatletter 
\setcounter{equation}{0} 
\setcounter{figure}{0} 
\renewcommand{\theequation}{E\arabic{equation}}
\renewcommand{\thefigure}{E\@arabic\c@figure}
 \makeatother

\section*{Appendix E: Model generalizations}

To illustrate the potential of the LA for investigating more sophisticated 
networks, here we provide the low-dimensional FREs corresponding to an 
excitatory network of QIF neurons with independently distributed
currents and synaptic weights, and to two 
interacting populations of excitatory and inhibitory QIF neurons
with distributed currents.

\subsection*{Excitatory population with heterogeneous currents
and synaptic weights}

\begin{figure}
\centerline{\includegraphics[width=65mm,clip=true]{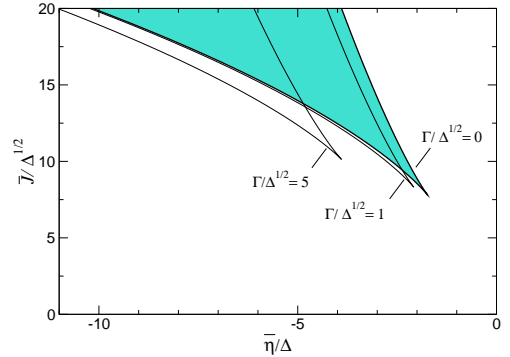}}
\caption{(color online). Phase diagram for an excitatory population with heterogeneous currents
and synaptic weights obtained using the FREs~\eqref{fre2}.}
\label{FigA5}
\end{figure}

As a first example, let us assume that {\em both} the currents $\eta$ and 
the synaptic weights $J$ are distributed ---this type of heterogeneity was also 
considered in \cite{bc13}. The input currents read then  
\footnote{Note that considering stronger forms of heterogeneity for 
the synaptic weights may generally not lead to Eq.~\eqref{general_QIF}.}
$$
I_j=\eta_j+J_j r(t) + I(t) .
$$
For simplicity, let us additionally assume that $\eta$ and $J$ 
are distributed independently, with a joint distribution
$p(\eta,J)=g(\eta) h(J)$. 
In the simplest situation of Lorentzian $g(\eta)$ and $h(J)$
\begin{equation}
g(\eta)= \frac{\Delta/\pi}{(\eta-\bar \eta)^2+\Delta^2} \,; \qquad h(J)=
\frac{\Gamma/\pi}{(J-\bar J)^2+\Gamma^2}.
\nonumber
\end{equation}
the problem is extremely simplified using the Lorentzian ansatz, which now trivially reads:
\begin{equation}
\rho(V|\eta,J,t)=\frac{1}{\pi}\frac{x(\eta,J,t)}{[V-y(\eta,J,t)]^2 + x(\eta,J,t)^2}. \nonumber
\end{equation}
The firing rate and mean membrane potential
are determined only by the value of $w\equiv x+iy$ at the poles in the lower half planes:
\begin{eqnarray}
r(t)+i v(t)&=& \iint w(\eta,J,t) g(\eta) h(J) d\eta dJ   \nonumber \\
&=& w(\bar\eta-i\Delta,\bar J-i\Gamma,t).  \nonumber
\end{eqnarray}
Finally, evaluating Eq.~\eqref{w} in the main text at the poles
($\eta=\bar\eta-i\Delta$, $J=\bar J - i\Gamma$) we get the exact FREs:
\begin{subequations}
\label{fre2}
\begin{eqnarray}
\dot r &=& \Delta/\pi + \Gamma r/\pi  + 2  r v, \label{freaa}\\ 
\dot v &=&   v^2 +   \bar \eta + \bar J r + I(t) - \pi^2 r^2  . \label{frebb}
\end{eqnarray}
\end{subequations}
These equations contain simply an extra term `$+\Gamma r/\pi$' compared 
to equations~\eqref{fre}.
Figure~\ref{FigA5} shows the bistability boundaries obtained from \eqref{fre2}
for different ratios of  $\Gamma$ and $\Delta$, and $I(t)=0$. Note that 
the region of bistability shifts to 
lower values of $\bar \eta/\Delta$ and to higher values of 
 $\bar J/\sqrt{\Delta}$ as the level of heterogeneity in the synaptic coupling
$\Gamma$ is increased.

\subsection*{Firing-rate equations for a pair of  
Excitatory-Inhibitory populations}

The microscopic state of each population of QIF neurons is characterized by the 
membrane potentials $\{V_j^{(e,i)}\}_{j=1,\ldots,N}$, which obey the following
ordinary differential equations:   
\begin{equation}
\dot V_j^{(e,i)}= (V_j^{(e,i)})^2+ I_j^{(e,i)}, \text{if } V_j^{(e,i)} \geq V_p, 
\text{then } V_j^{(e,i)} \leftarrow  V_r.
\nonumber
\end{equation}
Here, $V_j^{(e,i)}$ represents the membrane potential of neuron $j$ in either 
the excitatory $(e)$, or in the inhibitory population $(i)$. The external currents 
for the excitatory and inhibitory populations are, respectively:
\begin{eqnarray}
I_j^{(e)}= \eta_{j}^{(e)}+J_{ee} s^{(e)}(t)- J_{ie} s^{(i)}(t) +I^{(e)}(t), \nonumber \\
I_j^{(i)}=\eta_{j}^{(i)}+J_{ei} s^{(e)}(t)- J_{ii} s^{(i)}(t) +I^{(i)}(t), \nonumber
\end{eqnarray}
where the synaptic weights are $J_{ee},J_{ii},J_{ie},J_{ei}$.
Finally, the mean synaptic activation for each population is 
\begin{equation}
s^{(e,i)}(t)= \frac{1}{N} \sum_{j=1}^N  \sum_{k \backslash (t_j^k)^{(e,i)}<t}
\int_{-\infty}^{t}dt' a_\tau(t-t')\delta (t'-(t_j^{k})^{(e,i)}).\nonumber
\end{equation}
Here, $(t_j^k)^{(e,i)}$ is the time of the $k$th spike of $j$th neuron in 
either the excitatory $(e)$, or in the inhibitory population $(i)$. 
Additionally, $\delta (t)$ is the Dirac delta function, and $a_\tau(t)$
is the normalized synaptic activation caused by a single pre-synaptic 
spike with time scale $\tau$, e.g.~$a_\tau(t)=e^{-t/\tau}/\tau$.  

It is straightforward to apply the LA and the method described in the main text, 
to obtain the firing rate equations corresponding to the two-population model. 
Considering the limit of infinitely fast synapses, $\tau\to 0$, we get 
$s^{(e,i)}(t)=r^{(e,i)}(t)$. Finally, assuming the Lorentzian distributions 
of currents for both populations:
\begin{equation}
g_{e,i}(\eta)= \frac{1}{\pi} \frac{\Delta_{e,i}}{ (\eta-\bar\eta_{e,i})^2 +\Delta_{e,i}^2 },
\nonumber
\end{equation}
we obtain the firing-rate equations: 
\begin{eqnarray}
\dot r^{(e)} &=&  \Delta_e/\pi+ 2  r^{(e)} v^{(e)}, \nonumber \\
\dot v^{(e)} &=&  (v^{(e)})^2 + \bar \eta_e + J_{ee} r^{(e)}- J_{ie} r^{(i)} +I^{(e)}(t)
- (\pi r^{(e)})^2 , \nonumber \\
\dot r^{(i)} &=& \Delta_i/\pi+  2  r^{(i)} v^{(i)}, \nonumber \\
\dot v^{(i)} &=&  (v^{(i)})^2 +\bar \eta_i 
+ J_{ei} r^{(e)}- J_{ii} r^{(i)}+I^{(i)}(t) - (\pi r^{(i)})^2.\nonumber 
\label{ode2}
\end{eqnarray}

\end{document}